\documentclass[  superscriptaddress, amsmath,amssymb, aps,twocolumn]{revtex4-1}
\usepackage{graphicx}% Include figure files
\usepackage{dcolumn}% Align table columns on decimal point
\usepackage{amsthm}
\usepackage{bm}% bold math

\pdfoutput=1
\usepackage{graphicx,graphics,epsfig,subfigure,times,bm,bbm,amssymb,amsmath,amsthm,mathrsfs,MnSymbol}
\usepackage{gensymb}
\usepackage{amsfonts}
\usepackage[matrix,frame,arrow]{xypic}
\usepackage[pdfstartview=FitH]{hyperref}
\hypersetup{
	colorlinks=true,       % false: boxed links; true: colored links
	linkcolor=red,          % color of internal links
	citecolor=magenta,        % color of links to bibliography
	filecolor=magenta,      % color of file links
	urlcolor=cyan,           % color of external links
	runcolor=cyan}
\usepackage{epstopdf}
\usepackage[pdftex]{color}
\usepackage{braket}%Dirac Notation in QM
\usepackage{enumerate}
\usepackage[normalem]{ulem}
\usepackage[usenames,dvipsnames]{xcolor}
\usepackage{multirow}
\usepackage{mathtools}

\definecolor{orange}{rgb}{1,0.5,0}

\newcommand{\ignore}[1]{}
\usepackage{geometry}

\ignore{
	\documentclass[eprintnumbers,amsmath,amssymb,onecolumn,a4paper,caption ]{article}
	\usepackage{amsfonts}
	\usepackage{amssymb}
	\usepackage{mathrsfs}
	\usepackage{mathbbold}
	\usepackage{bbm}
	\usepackage{mathrsfs}
	\usepackage{dcolumn}% Align table columns on decimal point
	\usepackage{bm}% bold math
	\usepackage{times,epsfig,amssymb,amsmath}
	\usepackage{float}
	\usepackage{subfigure}
	\usepackage{geometry}\geometry{left=2.5cm,right=2.5cm,top=3cm,bottom=3cm}

	\usepackage{color}

}

\begin{document}

\title{Observation of strong and weak thermalization in a superconducting quantum processor}

\author{Fusheng Chen}
\thanks{Those authors contributed equally to this work.}
\affiliation{Hefei National Laboratory for Physical Sciences at Microscale and Department of Modern Physics, University of Science and Technology of China, Hefei, Anhui 230026, China}
\affiliation{Shanghai Branch, CAS Center for Excellence and Synergetic Innovation Center in Quantum Information and Quantum Physics, University of Science and Technology of China, Shanghai 201315, China}
\affiliation{Shanghai Research Center for Quantum Sciences, Shanghai 201315, China}

\author{Zheng-Hang Sun}
\thanks{Those authors contributed equally to this work.}
\affiliation{Institute of Physics, Chinese Academy of Sciences, Beijing 100190, China}
\affiliation{School of Physical Sciences, University of Chinese Academy of Sciences, Beijing 100190, China}

\author{Ming Gong}
\thanks{Those authors contributed equally to this work.}
\affiliation{Hefei National Laboratory for Physical Sciences at Microscale and Department of Modern Physics, University of Science and Technology of China, Hefei, Anhui 230026, China}
\affiliation{Shanghai Branch, CAS Center for Excellence and Synergetic Innovation Center in Quantum Information and Quantum Physics, University of Science and Technology of China, Shanghai 201315, China}
\affiliation{Shanghai Research Center for Quantum Sciences, Shanghai 201315, China}

\author{Qingling Zhu}
\affiliation{Hefei National Laboratory for Physical Sciences at Microscale and Department of Modern Physics, University of Science and Technology of China, Hefei, Anhui 230026, China}
\affiliation{Shanghai Branch, CAS Center for Excellence and Synergetic Innovation Center in Quantum Information and Quantum Physics, University of Science and Technology of China, Shanghai 201315, China}
\affiliation{Shanghai Research Center for Quantum Sciences, Shanghai 201315, China}

\author{Yu-Ran Zhang}
\affiliation{Theoretical Quantum Physics Laboratory, RIKEN Cluster for Pioneering Research, Wako-shi, Saitama 351-0198, Japan}

\author{Yulin Wu}
\affiliation{Hefei National Laboratory for Physical Sciences at Microscale and Department of Modern Physics, University of Science and Technology of China, Hefei, Anhui 230026, China}
\affiliation{Shanghai Branch, CAS Center for Excellence and Synergetic Innovation Center in Quantum Information and Quantum Physics, University of Science and Technology of China, Shanghai 201315, China}
\affiliation{Shanghai Research Center for Quantum Sciences, Shanghai 201315, China}

\author{Yangsen Ye}
\affiliation{Hefei National Laboratory for Physical Sciences at Microscale and Department of Modern Physics, University of Science and Technology of China, Hefei, Anhui 230026, China}
\affiliation{Shanghai Branch, CAS Center for Excellence and Synergetic Innovation Center in Quantum Information and Quantum Physics, University of Science and Technology of China, Shanghai 201315, China}
\affiliation{Shanghai Research Center for Quantum Sciences, Shanghai 201315, China}

\author{Chen Zha}
\affiliation{Hefei National Laboratory for Physical Sciences at Microscale and Department of Modern Physics, University of Science and Technology of China, Hefei, Anhui 230026, China}
\affiliation{Shanghai Branch, CAS Center for Excellence and Synergetic Innovation Center in Quantum Information and Quantum Physics, University of Science and Technology of China, Shanghai 201315, China}
\affiliation{Shanghai Research Center for Quantum Sciences, Shanghai 201315, China}

\author{Shaowei Li}
\affiliation{Hefei National Laboratory for Physical Sciences at Microscale and Department of Modern Physics, University of Science and Technology of China, Hefei, Anhui 230026, China}
\affiliation{Shanghai Branch, CAS Center for Excellence and Synergetic Innovation Center in Quantum Information and Quantum Physics, University of Science and Technology of China, Shanghai 201315, China}
\affiliation{Shanghai Research Center for Quantum Sciences, Shanghai 201315, China}

\author{Shaojun Guo}
\affiliation{Hefei National Laboratory for Physical Sciences at Microscale and Department of Modern Physics, University of Science and Technology of China, Hefei, Anhui 230026, China}
\affiliation{Shanghai Branch, CAS Center for Excellence and Synergetic Innovation Center in Quantum Information and Quantum Physics, University of Science and Technology of China, Shanghai 201315, China}
\affiliation{Shanghai Research Center for Quantum Sciences, Shanghai 201315, China}

\author{Haoran Qian}
\affiliation{Hefei National Laboratory for Physical Sciences at Microscale and Department of Modern Physics, University of Science and Technology of China, Hefei, Anhui 230026, China}
\affiliation{Shanghai Branch, CAS Center for Excellence and Synergetic Innovation Center in Quantum Information and Quantum Physics, University of Science and Technology of China, Shanghai 201315, China}
\affiliation{Shanghai Research Center for Quantum Sciences, Shanghai 201315, China}

\author{He-Liang Huang}
\affiliation{Hefei National Laboratory for Physical Sciences at Microscale and Department of Modern Physics, University of Science and Technology of China, Hefei, Anhui 230026, China}
\affiliation{Shanghai Branch, CAS Center for Excellence and Synergetic Innovation Center in Quantum Information and Quantum Physics, University of Science and Technology of China, Shanghai 201315, China}
\affiliation{Shanghai Research Center for Quantum Sciences, Shanghai 201315, China}

\author{Jiale Yu}
\affiliation{Hefei National Laboratory for Physical Sciences at Microscale and Department of Modern Physics, University of Science and Technology of China, Hefei, Anhui 230026, China}
\affiliation{Shanghai Branch, CAS Center for Excellence and Synergetic Innovation Center in Quantum Information and Quantum Physics, University of Science and Technology of China, Shanghai 201315, China}
\affiliation{Shanghai Research Center for Quantum Sciences, Shanghai 201315, China}

\author{Hui Deng}
\affiliation{Hefei National Laboratory for Physical Sciences at Microscale and Department of Modern Physics, University of Science and Technology of China, Hefei, Anhui 230026, China}
\affiliation{Shanghai Branch, CAS Center for Excellence and Synergetic Innovation Center in Quantum Information and Quantum Physics, University of Science and Technology of China, Shanghai 201315, China}
\affiliation{Shanghai Research Center for Quantum Sciences, Shanghai 201315, China}

\author{Hao Rong}
\affiliation{Hefei National Laboratory for Physical Sciences at Microscale and Department of Modern Physics, University of Science and Technology of China, Hefei, Anhui 230026, China}
\affiliation{Shanghai Branch, CAS Center for Excellence and Synergetic Innovation Center in Quantum Information and Quantum Physics, University of Science and Technology of China, Shanghai 201315, China}
\affiliation{Shanghai Research Center for Quantum Sciences, Shanghai 201315, China}

\author{Jin Lin}
\affiliation{Hefei National Laboratory for Physical Sciences at Microscale and Department of Modern Physics, University of Science and Technology of China, Hefei, Anhui 230026, China}
\affiliation{Shanghai Branch, CAS Center for Excellence and Synergetic Innovation Center in Quantum Information and Quantum Physics, University of Science and Technology of China, Shanghai 201315, China}
\affiliation{Shanghai Research Center for Quantum Sciences, Shanghai 201315, China}

\author{Yu Xu}
\affiliation{Hefei National Laboratory for Physical Sciences at Microscale and Department of Modern Physics, University of Science and Technology of China, Hefei, Anhui 230026, China}
\affiliation{Shanghai Branch, CAS Center for Excellence and Synergetic Innovation Center in Quantum Information and Quantum Physics, University of Science and Technology of China, Shanghai 201315, China}
\affiliation{Shanghai Research Center for Quantum Sciences, Shanghai 201315, China}

\author{Lihua Sun}
\affiliation{Hefei National Laboratory for Physical Sciences at Microscale and Department of Modern Physics, University of Science and Technology of China, Hefei, Anhui 230026, China}
\affiliation{Shanghai Branch, CAS Center for Excellence and Synergetic Innovation Center in Quantum Information and Quantum Physics, University of Science and Technology of China, Shanghai 201315, China}
\affiliation{Shanghai Research Center for Quantum Sciences, Shanghai 201315, China}

\author{Cheng Guo}
\affiliation{Hefei National Laboratory for Physical Sciences at Microscale and Department of Modern Physics, University of Science and Technology of China, Hefei, Anhui 230026, China}
\affiliation{Shanghai Branch, CAS Center for Excellence and Synergetic Innovation Center in Quantum Information and Quantum Physics, University of Science and Technology of China, Shanghai 201315, China}
\affiliation{Shanghai Research Center for Quantum Sciences, Shanghai 201315, China}

\author{Na Li}
\affiliation{Hefei National Laboratory for Physical Sciences at Microscale and Department of Modern Physics, University of Science and Technology of China, Hefei, Anhui 230026, China}
\affiliation{Shanghai Branch, CAS Center for Excellence and Synergetic Innovation Center in Quantum Information and Quantum Physics, University of Science and Technology of China, Shanghai 201315, China}
\affiliation{Shanghai Research Center for Quantum Sciences, Shanghai 201315, China}

\author{Futian Liang}
\affiliation{Hefei National Laboratory for Physical Sciences at Microscale and Department of Modern Physics, University of Science and Technology of China, Hefei, Anhui 230026, China}
\affiliation{Shanghai Branch, CAS Center for Excellence and Synergetic Innovation Center in Quantum Information and Quantum Physics, University of Science and Technology of China, Shanghai 201315, China}
\affiliation{Shanghai Research Center for Quantum Sciences, Shanghai 201315, China}

\author{Cheng-Zhi Peng}
\affiliation{Hefei National Laboratory for Physical Sciences at Microscale and Department of Modern Physics, University of Science and Technology of China, Hefei, Anhui 230026, China}
\affiliation{Shanghai Branch, CAS Center for Excellence and Synergetic Innovation Center in Quantum Information and Quantum Physics, University of Science and Technology of China, Shanghai 201315, China}
\affiliation{Shanghai Research Center for Quantum Sciences, Shanghai 201315, China}

\author{Heng Fan}
%\email{hfan@iphy.ac.cn}
\affiliation{Institute of Physics, Chinese Academy of Sciences, Beijing 100190, China}
\affiliation{School of Physical Sciences, University of Chinese Academy of Sciences, Beijing 100190, China}
\affiliation{Songshan Lake  Materials Laboratory, Dongguan 523808, Guangdong, China}
\affiliation{CAS Center for Excellent in Topological Quantum Computation, University of Chinese Academy of Sciences, Beijing 100190, China}

\author{Xiaobo Zhu}
%\email{xbzhu16@ustc.edu.cn}
\affiliation{Hefei National Laboratory for Physical Sciences at Microscale and Department of Modern Physics, University of Science and Technology of China, Hefei, Anhui 230026, China}
\affiliation{Shanghai Branch, CAS Center for Excellence and Synergetic Innovation Center in Quantum Information and Quantum Physics, University of Science and Technology of China, Shanghai 201315, China}
\affiliation{Shanghai Research Center for Quantum Sciences, Shanghai 201315, China}

\author{Jian-Wei Pan}
\affiliation{Hefei National Laboratory for Physical Sciences at Microscale and Department of Modern Physics, University of Science and Technology of China, Hefei, Anhui 230026, China}
\affiliation{Shanghai Branch, CAS Center for Excellence and Synergetic Innovation Center in Quantum Information and Quantum Physics, University of Science and Technology of China, Shanghai 201315, China}
\affiliation{Shanghai Research Center for Quantum Sciences, Shanghai 201315, China}

\begin{abstract}
\noindent We experimentally study the ergodic dynamics of a 1D array of 12 superconducting qubits with a transverse field, and identify the regimes of strong and weak thermalization with different initial states. We observe convergence of the local observable to its thermal expectation value in the strong-thermalizaion regime. For weak thermalization, the dynamics of local observable exhibits an oscillation around the thermal value, which can only be attained by the time average. We also demonstrate that the entanglement entropy and concurrence can characterize the regimes of strong and weak thermalization. Our work provides an essential step towards a generic understanding of thermalization in quantum systems.
\end{abstract}
\pacs{Valid PACS appear here}
\maketitle

Statistical mechanics is developed to describe the thermodynamics of both classical and quantum systems. If a balloon, containing a large number of air molecules, is pierced in an evacuated chamber, the molecules will move around all possible states, and with the long-time average, they will satisfy the Maxwell velocity distribution, which is independent of the initial condition~\cite{thermal_book}. In quantum cases, thermal equilibrium states, described by the statistical-mechanical prescription, are expected to emerge in the non-equilibrium dynamics of a closed non-integrable many-body system~\cite{thermal1,thermal2,thermal3,thermal4,thermal5,thermal6}. Different from the classical case, the time average is not necessary for quantum thermalization, and the quenched states could converge to their thermal expectations at a time after a short relaxation~\cite{thermal2}. This phenomenon is regarded as strong thermalization. However, strong thermalization cannot be certainly achieved by a non-integrable system driven out of equilibrium, whose occurrence is also relevant to the choice of initial states. Numerical works~\cite{strong_weak_1,strong_weak_2} have revealed that in a non-integrable 1D Ising model, strong thermalization happens when the effective inverse temperature of initial states is close to 0. In contrast, if the effective inverse temperature of initial states is sufficiently far away from 0, the temporal evolution of the local observable shows an obvious oscillation, with the long-time average attaining the thermal expectation value. This phenomenon is known as weak thermalization. Recently, it has been numerically shown that regimes of strong and weak thermalization exist in the long-range Ising model describing trapped ions~\cite{strong_weak_3}. Nevertheless, the experimental observation of both strong and weak thermalization remains absent.

%\textbf{Superconducting quantum processor and experimental pulse sequence for characterizing strong and weak thermalization with different initial states.}
\begin{figure*}[]
	\centering
	\includegraphics[width=1\linewidth]{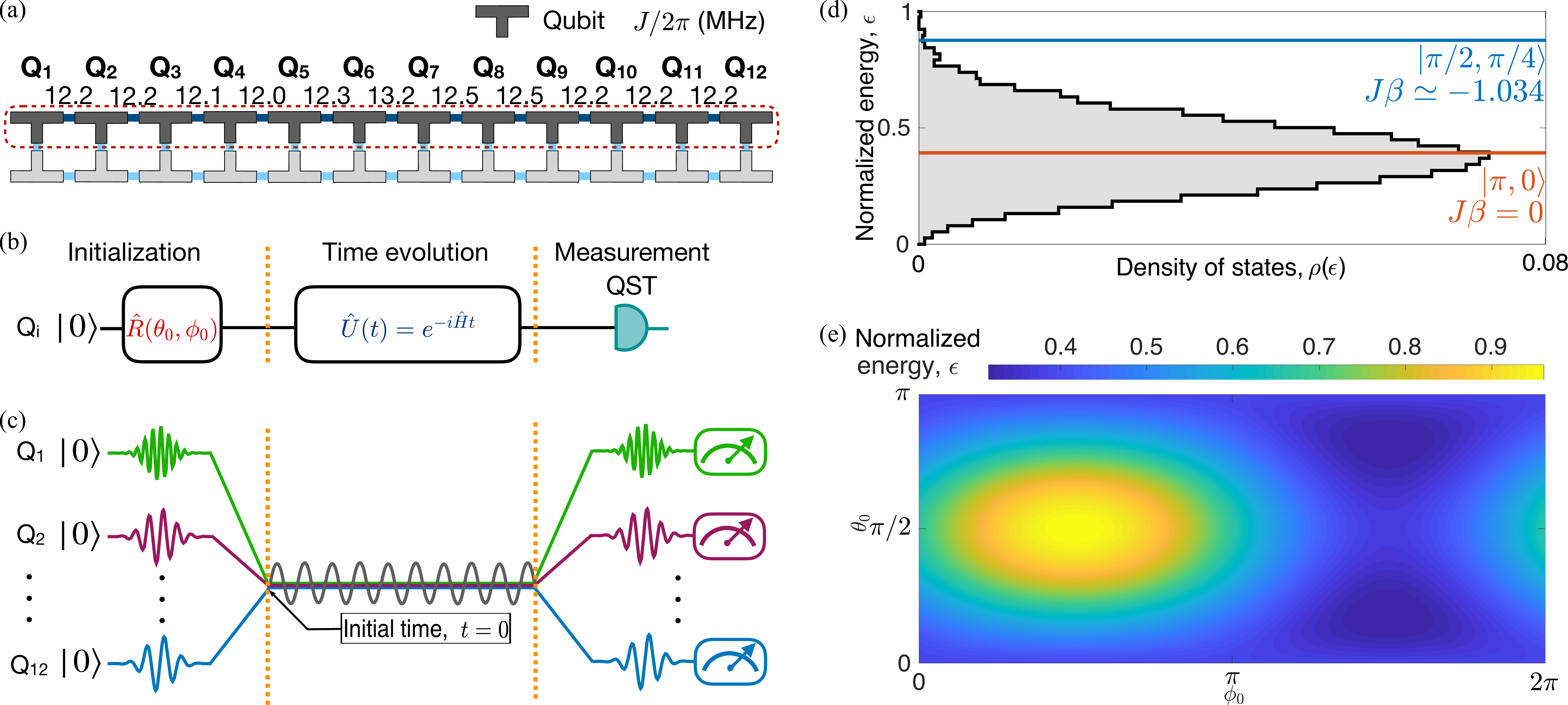}\\
	\caption{(a) Architecture of the 24-qubit superconducting circuit. The qubits $Q_{1}$-$Q_{12}$ are employed to realize a 1D non-integrable model. (b) The schematic diagram of the pulse sequence used to observe strong and weak thermalization, which consists of three parts, i.e., initialization, evolution and measurement. For the initialization, we prepare all qubits at $|\theta_{0},\phi_{0}\rangle$, i.e., Eq.~(\ref{initial}). Starting from the $|0\rangle$ state, all qubits are prepared at $|\theta_{0},\phi_{0}'\rangle$ via the gate $\hat{R}(\theta_{0},\phi_{0}')$ rotating $|0\rangle$ around the axis $\hat{n} = \cos(\phi_{0}') \hat{\sigma}^{y} - \sin(\phi_{0}') \hat{\sigma}^{x}$ by a angle $\theta_{0}$. However, due to the dynamical phases accumulated in tunning qubits to the evolution point, $\phi_{0}'$ is not equal to $\phi_{0}$. We calibrate the dynamical phases, and then correct all qubits to the same initial state $|\theta_{0},\phi_{0}\rangle$. (c) The control waveforms corresponding to the pulse sequence in (b). The single-qubit pulses after $|0\rangle$ refer to the gate $\hat{R}(\theta_{0},\phi_{0})$. To realize the non-equilibrium quantum dynamics, the Z pulses and resonant microwave pulses (the sinusoidal line) are simultaneously applied to each qubit. After the evolution, quantum state tomography measurements are performed at idle points of the qubits and single-qubit pulses are required. (d) For the Hamiltonian (\ref{XY_Y}), the density of states $\rho(\epsilon)$ as a function of the normalized energy $\epsilon$ obtained by the numerical simulation. The normalized energy of two initial states $|\pi,0\rangle$ and $|\pi/2,\pi/4\rangle$ are highlighted. (e) The normalized energy of the initial state $|\theta_{0},\phi_{0}\rangle$ as a function of $\theta_{0}$ and $\phi_{0}$.  }\label{exp_setup}
\end{figure*}

On the basis of the high-precision control, long coherence time, and the accurate readout, a superconducting quantum processor is an excellent platform for generating multipartite entangled states~\cite{GHZ1,GHZ2,GHZ3}, characterizing quantum supremacy~\cite{qs1,qs2,qs3}, and demonstrating variational quantum computation~\cite{vqe1,vqe2}. Moreover, by performing analog quantum simulations, the platform is also employed to study the phenomena in quantum many-body systems out of equilibrium, including quantum walks~\cite{qw}, many-body localization~\cite{mbme,mbl_a1,mbl_a2}, dynamical phase transitions~\cite{sciadv}, and ergodic-localized junctions~\cite{elj}.

Here, we realize a non-integrable system using a 1D array of 12 superconducting qubits with a controllable transverse field. We experimentally observe the signatures of strong and weak thermalization via measuring the local observable with different initial states. Since the description of the local observable, using statistical mechanics, relies on the local entropy created by entanglement, the dynamics of the entanglement entropy (EE) plays a key role in thermalization~\cite{EE1,EE2,EE3,EE4,EE5}. Thus, we study the EE of the single-qubit subsystems and show that the EE can distinguish the strong-thermalization regime from the weak one. Furthermore, we measure the concurrence~\cite{concurrence1} of the reduced density matrices of two nearest qubits, employing the tomographic readout, and observe thermal entanglement~\cite{concurrence2} in the presence of weak thermalization.

The experiment is performed on a chain of 12 superconducting transmon qubits [see Fig.~\ref{exp_setup}(a)], described by the Hamiltonian of the Bose-Hubbard model~\cite{qw,BH_a1,BH1,BH3}
\begin{eqnarray} \nonumber
\hat{H} &=& J\sum_{j=1}^{11} (\hat{a}_{j}^{\dagger}\hat{a}_{j+1}+ \text{H.c.}) \\
&+&  \frac{U}{2}\sum_{j=1}^{12}\hat{n}_{j}(\hat{n}_{j}-1) +\sum_{j=1}^{12}\mu_{j}\hat{n}_{j},
\label{SQP}
\end{eqnarray}
where $\hat{n}_{j}=\hat{a}_{j}^{\dagger}\hat{a}_{j}$ is the bosonic number operator, with $\hat{a}_{j}^{\dagger}$ ($\hat{a}_{j}$) being the creation (annihilation) operator, and $J$, $U$, and $\mu_{n}$ denote the nearest-neighbor coupling strength, the on-site nonlinear interaction, and tunable on-site potential, respectively. More details of our device can be found in Supplementary Materials~\cite{SM}.

With $U/J\rightarrow\infty$ and all $\mu_{n}$ being tuned to the same frequency, the Hamiltonian (\ref{SQP}) can be rewritten as~\cite{BH4,BH5}
\begin{eqnarray}
\hat{H} &=& \lambda \sum_{j=1}^{11} (\hat{\sigma}_{j}^{x}\hat{\sigma}_{j+1}^{x} + \hat{\sigma}_{j}^{y}\hat{\sigma}_{j+1}^{y}),
\label{XY}
\end{eqnarray}
with $\lambda = J/2$ and $\hat{\sigma}_{j}^{\alpha} \indent (\alpha\in\{x,y,z\})$ being the Pauli matrices. The Hamiltonian (\ref{XY}) describes an integrable model, which can be exactly solved via introducing the Jordan-Wigner transformation~\cite{XY_chain}. To realize a non-integrable system where thermalization occurs, we impose resonant microwave drives with a magnitude $g\simeq\lambda$ on all qubits, generating a local transverse field~\cite{sciadv}. The final effective Hamiltonian of the system is
\begin{eqnarray}
\hat{H} &=& \lambda \sum_{j=1}^{11} (\hat{\sigma}_{j}^{x}\hat{\sigma}_{j+1}^{x} + \hat{\sigma}_{j}^{y}\hat{\sigma}_{j+1}^{y}) + g\sum_{j=1}^{12} \hat{\sigma}_{j}^{y}.
\label{XY_Y}
\end{eqnarray}
For details regarding the Hamiltonian (\ref{XY_Y}), see Supplementary Materials~\cite{SM}.

%\textbf{The local observable and entanglement entropy. }
\begin{figure*}[]
	\centering
	\includegraphics[width=1\linewidth]{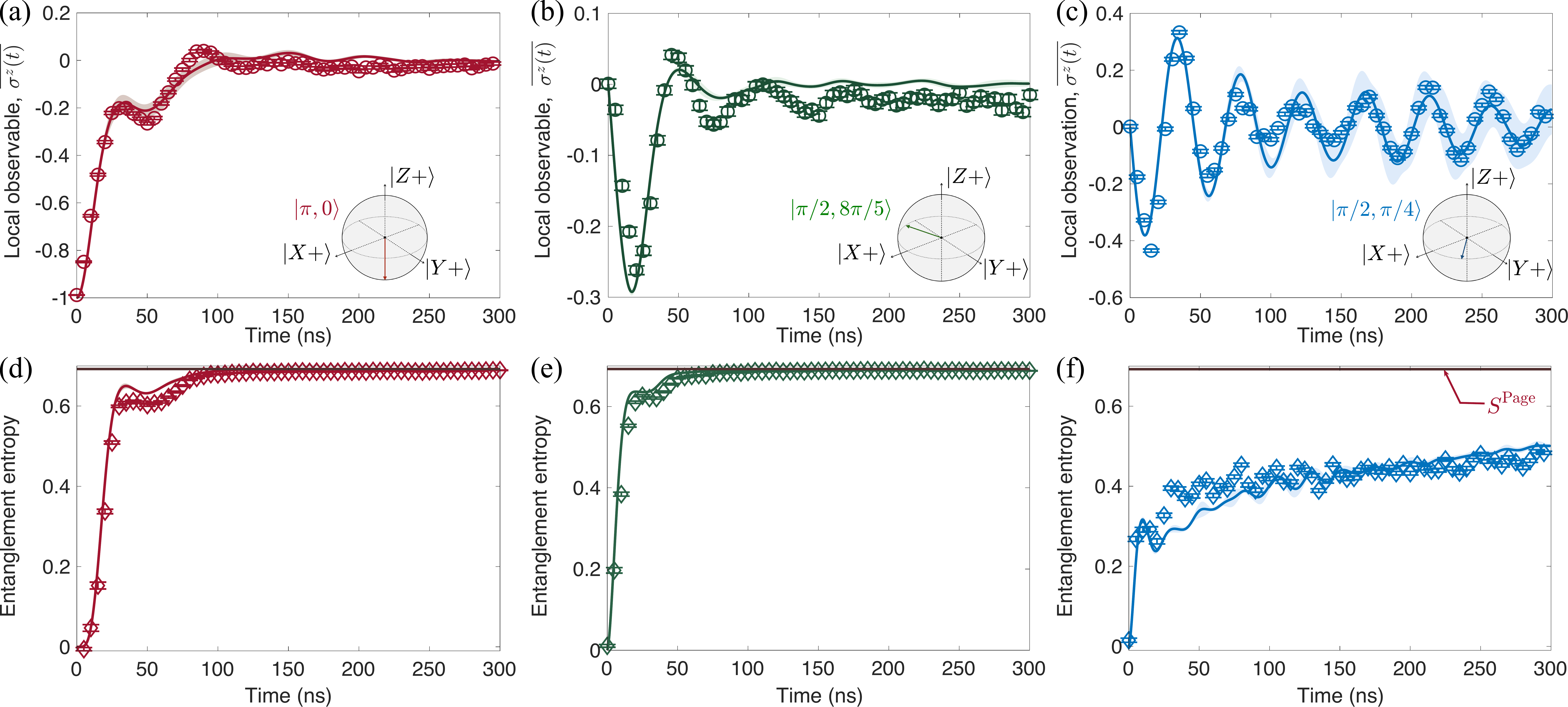}\\
	\caption{(a) Experimental data of the time evolution of the local observable $\overline{\hat{\sigma}^{z}(t)}$ with the initial state $|\pi,0\rangle$. (b) As in (a), but for the initial state $|\pi/2,8\pi/5\rangle$. (c) As in (a), but for the initial state $|\pi/2,\pi/4\rangle$. (d) Experimental data of the time evolution of the entanglement entropy with the initial state $|\pi,0\rangle$. (e) As in (d), but for the initial state $|\pi/2,8\pi/5\rangle$. (f) As in (d), but for the initial state $|\pi/2,\pi/4\rangle$. The horizontal lines in (d)-(f) denote the Page value of the EE. The solid lines in (a)-(f) are numerical results without considering decoherence. The solid line in (f) is numerical results considering decoherence (see Supplementary Materials~\cite{SM} for the effects of decoherence). The shaded region shows the errorbars of the numerical results, taking the uncertainties of the local field into consideration (see Supplementary Materials~\cite{SM}). The initial states are presented in Bloch spheres, where $|X+\rangle$, $|Y+\rangle$, and $|Z+\rangle$ are the eigenstate of $\sigma^{x}$, $\sigma^{y}$ and $\sigma^{z}$ with the eigenvalue $+1$, respectively. }\label{fig2}
\end{figure*}

%\textbf{Time average of the entanglement entropy. }
\begin{figure}[]
	\centering
	\includegraphics[width=1\linewidth]{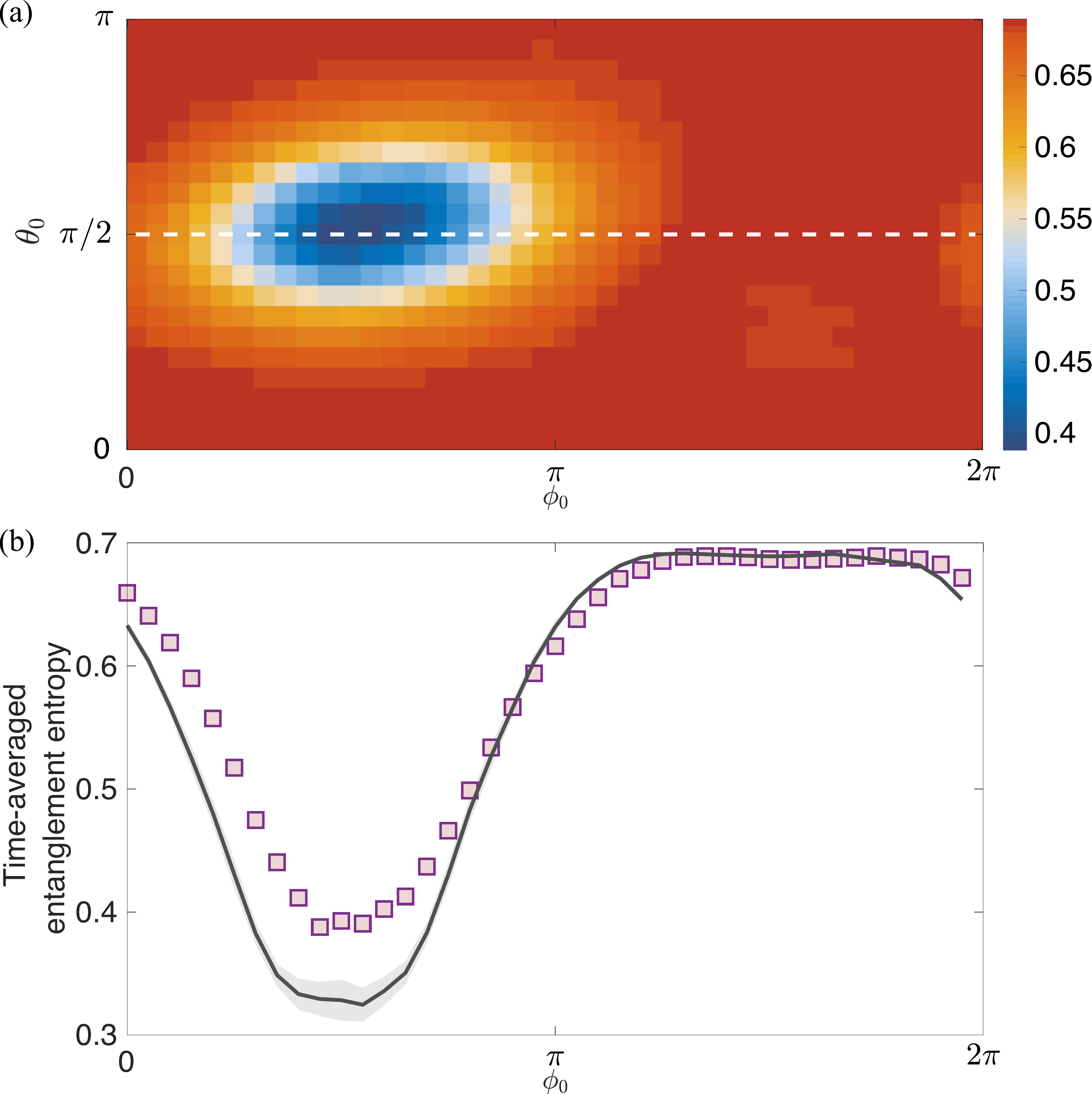}\\
	\caption{ (a) Experimental data of the time-averaged EE with initial states $|\theta_{0},\phi_{0}\rangle$ plotted in the $\phi_{0}-\theta_{0}$ plane. (b) Experimental data of the time-averaged EE with initial states $|\pi/2,\phi_{0}\rangle$. The white dashed line in (a) highlights $\theta_{0}=\pi/2$. The solid line in (b) shows the numerical result considering decoherence (see Supplementary Materials~\cite{SM} for the effects of decoherence). The shaded region shows the errorbars of the numerical results, taking the uncertainties of the local field into consideration (see Supplementary Materials~\cite{SM}).}\label{fig3}
\end{figure}

To observe strong and weak thermalization, we initialize the system by preparing each qubit in the direction $(\theta_{0},\phi_{0})$, which can be described as the spin coherent state
\begin{eqnarray}
|\theta_{0},\phi_{0}\rangle = \prod_{j=1}^{12} (\cos\frac{\theta_{0}}{2}|+Z\rangle_{j} + e^{-i\phi_{0}}\sin\frac{\theta_{0}}{2}|-Z\rangle_{j}),
\label{initial}
\end{eqnarray}
where $|+Z\rangle_{j}$ ($|-Z\rangle_{j}$) denotes the eigenstate of $\hat{\sigma}^{z}_{j}$ with the eigenvalue $+1$ ($-1$). Next, all qubits are biased to the working point to start the quench dynamics, $|\Psi_{t}\rangle = e^{-i\hat{H}t}|\theta_{0},\phi_{0}\rangle$. We then tune the qubits to their idle points, and perform the quantum state tomography to reconstruct the one- and two-qubit density matrices. The experimental pulse sequence and control waveforms are shown in Fig.~\ref{exp_setup}(b) and (c), respectively. There are three essential experimental requirements to be satisfied: ($i$) To realize the time evolution, all qubits should be tuned to the same frequency; ($ii$) The initial states of all qubits should be uniform at the start point of the time evolution; ($iii$) The local transverse fields of all qubits should be uniform during the evolution. These requirements are fulfilled with specific calibrations (see Supplementary Materials for details~\cite{SM}).

The occurrence of strong or weak thermalization relates closely to
the effective inverse temperature $\beta$ of $|\theta_{0},\phi_{0}\rangle$, which can be obtained by solving $\text{Tr}[\hat{H}(|\theta_{0},\phi_{0}\rangle\langle\theta_{0},\phi_{0}|-\hat{\rho}_{\beta})]=0$, with $\hat{\rho}_{\beta} =e^{-\beta\hat{H}}/\text{Tr}(e^{-\beta\hat{H}})$ being the thermal state~\cite{strong_weak_1}. Moreover, the quasiparticle explanation of weak thermalization indicates that initial states in the weak-thermalization regime are near the edge of the energy spectrum~\cite{strong_weak_2}. Here, we first consider two initial states $|\theta_{0},\phi_{0}\rangle = |\pi/2,\pi/4\rangle$ and $|\pi,0\rangle$ whose effective inverse temperature are numerically estimated as $J\beta\simeq -1.034$ and $0$, lying in the weak- and strong-thermalization regime, respectively. In addition to the effective inverse temperature, the regimes of strong and weak thermalization can be identified by defining the normalized energy of the initial state $|\theta_{0},\phi_{0}\rangle$
\begin{eqnarray}
\epsilon= \frac{\langle\theta_{0},\phi_{0}|\hat{H}|\theta_{0},\phi_{0}\rangle -E_{\text{min}}}{E_{\text{max}} - E_{\text{min}}}
\label{ne}
\end{eqnarray}
with $E_{\text{max}}$ and $E_{\text{min}}$ being the maximum and minimum eigenvalue of $\hat{H}$, respectively. In Fig.~\ref{exp_setup}(d), the $\epsilon$ of the initial state $|\pi,0\rangle$ corresponds to the maximum density of states (DoS) $\rho(\epsilon)$, while the $\epsilon$ of the initial state $|\pi/2,\pi/4\rangle$ is close to the edge with $\rho(\epsilon)\simeq 0$. Moreover, in Fig.~\ref{exp_setup}(e), we plot the normalized energy of different initial states $|\theta_{0},\phi_{0}\rangle$, i.e., $\epsilon(\theta_{0},\phi_{0})$.

We start by characterizing strong and weak thermalization employing the local observable $\overline{\hat{\sigma}^{z}(t)} =\frac{1}{12}\sum_{j=1}^{12} \langle\Psi_{t} |\hat{\sigma}^{z}_{j}|\Psi_{t}\rangle $.
Figure~\ref{fig2}(a) and (c) present the experimental results of the time evolution of $\overline{\hat{\sigma}^{z}(t)}$ with initial states $|\pi,0\rangle$ and $|\pi/2,\pi/4\rangle$, respectively. It is shown that for the initial state $|\pi,0\rangle$ in the strong-thermalization regime, $\overline{\hat{\sigma}^{z}(t)}$ stably achieve the thermal value $\text{Tr}(\hat{\rho}_{\beta}\hat{\sigma}^{z}_{j})=0$ after $t\simeq 150$ ns. In contrast, for the initial state $|\pi/2,\pi/4\rangle$ in the weak-thermalization regime, $\overline{\hat{\sigma}^{z}(t)}$ strongly oscillates around the thermal value 0. In addition, we measure the dynamics of $\overline{\hat{\sigma}^{z}(t)}$ with the initial state $|\pi/2,8\pi/5\rangle$, which also lies in the strong-thermalization regime, since its effective inverse temperature is $J\beta\simeq 0$. The results, depicted in Fig.~\ref{fig2}(b), show that even the behavior of short relaxation is different from that with the initial state $|\pi,0\rangle$, the local observable also has a stationary value near the thermal value $0$ after $t\simeq 150$ ns, which is a hallmark of strong thermalization.

Next, we consider the von Neumann entanglement entropy (EE), $S=-\text{Tr}[\hat{\rho}_{j}\ln(\hat{\rho}_{j})]$, where $\hat{\rho}_{j}$ is the reduced density matrix of the $j$-th qubit. We average the EE over all qubit sites. The dynamics of the EE, with the initial states $|\pi,0\rangle$ and $|\pi/2,8\pi/5\rangle$ in the regime of strong thermalization, are displayed in Fig.~\ref{fig2}(d) and (e), respectively. We observe that for strong thermalization, the EE rapidly reaches the Page value $S^{\text{Page}}\simeq 0.692$ as the maximum EE of a single-qubit subsystem of a total system in the random pure state~\cite{Page}. However, for weak thermalization, the non-equilibrium dynamics gains the EE smaller than the Page value [Fig.~\ref{fig2}(f)].

Furthermore, we study the time-averaged EE between $100$ ns and $200$ ns with different initial states $|\theta_{0},\phi_{0}\rangle$. In Fig.~\ref{fig3}(a), we show the experimental data of time-averaged EE with different initial states $|\theta_{0},\phi_{0}\rangle$, which bears a close resemblance of to the normalized energy in Fig.~\ref{exp_setup}(e). Specifically, with $\theta_{0}=\pi/2$, around $\phi_{0}\simeq 3\pi/2$, $\epsilon\simeq 0.4$, and the DoS $\rho(\epsilon)$ becomes the maximum [see Fig.~\ref{exp_setup}(d) and (e)]. Thus, it can be predicted that strong thermalization occurs in this regime. Additionally, according to the results in Fig.~\ref{exp_setup}(d) and (e), the normalized energy is near 1 at $\phi_{0}=\pi/2$, where the DoS is close to 0, and the degree of thermalization is the weakest. The experimental data of the time-averaged EE, with $\theta_{0}=\pi/2$ and different $\phi_{0}$, are presented in Fig.~\ref{fig3}(b). There is a minimum of the EE around $\phi_{0}=\pi/2$, corresponding to the weakest thermalization. Moreover, the maximum EE reveals a regime of strong thermalization with $\theta_{0}=\pi/2$ and $\phi_{0}\in [1.3\pi,1.9\pi]$.

%\textbf{Strong and weak thermalization characterized by the trace distance and concurrence. }
\begin{figure}[]
	\centering
	\includegraphics[width=1\linewidth]{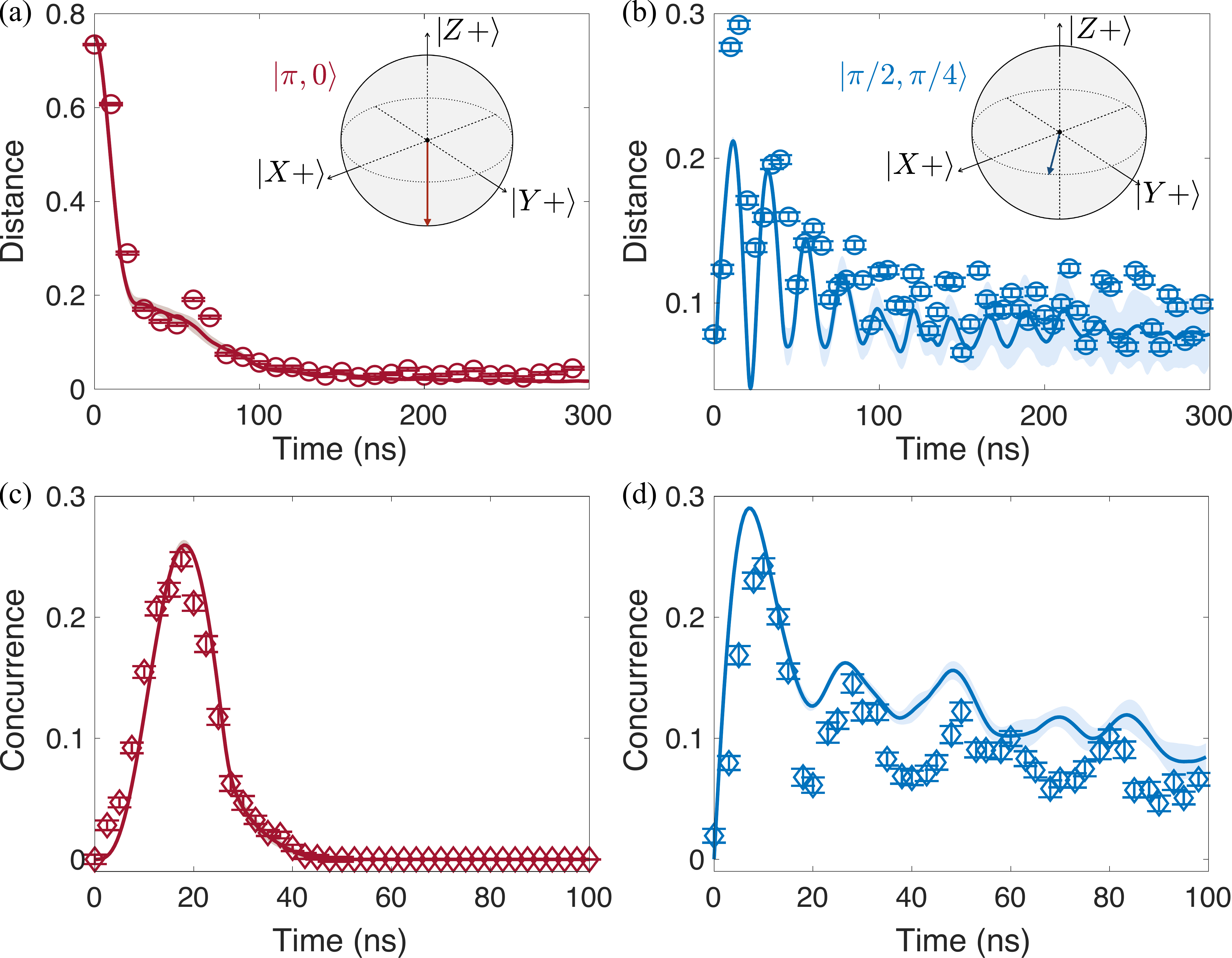}\\
	\caption{ (a) Experimental data of the trace distance between the non-equilibrium and thermal states with the initial state $|\pi,0\rangle$. (b) As in (a), but for the initial state $|\pi/2,\pi/4\rangle$. (c) Experimental data of the concurrence with the initial state $|\pi,0\rangle$. (d) As in (c), but for the initial state $|\pi/2,\pi/4\rangle$. The solid lines in (a) and (c) are numerical results without considering decoherence. The solid lines in (b) and (c) are numerical results considering decoherence (see Supplementary Materials~\cite{SM} for the effects of decoherence). The shaded region shows the errorbars of the numerical results, taking the uncertainties of the local field into consideration (see Supplementary Materials~\cite{SM}). The initial states are presented in Bloch spheres, where $|X+\rangle$, $|Y+\rangle$, and $|Z+\rangle$ are the eigenstate of $\sigma^{x}$, $\sigma^{y}$ and $\sigma^{z}$ with the eigenvalue $+1$, respectively.}\label{fig4}
\end{figure}

The trace distance between the non-equilibrium state $\hat{\rho}_{t} = |\Psi_{t}\rangle\langle\Psi_{t} |$ and the thermal state $\hat{\rho}_{\beta}$, with the $\beta$ being the effective inverse temperature of the initial state, i.e., $\frac{1}{2}\text{Tr}(|\hat{\rho}_{t} - \hat{\rho}_{\beta}|)$, can diagnose quantum thermalization~\cite{EE1}. It has been numerically shown that the distance monotonically decays to 0 for strong thermalization. With initial states in the weak-thermalization regime, the decay of the distance can also be observed but with a strong fluctuation~\cite{strong_weak_1}.

We measure the reduced density matrix $\hat{\rho}_{t}^{j,j+1}$ of the subsystem consisting of the $j$-th and $(j+1)$-th qubit, using the quantum state tomography. For the initial state $|\pi,0\rangle$, the effective inverse temperature is $J\beta=0$, and the corresponding two-qubit thermal state is $\hat{I}/4$ with $\hat{I}$ being an identity matrix. Then, the trace distance between $\hat{\rho}_{t}^{j,j+1}$ and $\hat{I}/4$, averaged over the qubit site $j$, can be directly obtained. Similarly, by considering the thermal states $\hat{\rho}_{\beta}^{j,j+1}$ with $J\beta\simeq -1.034$, the dynamics of the trace distance, with the initial state $|\pi/2,\pi/4\rangle$, can also be measured. As shown in Fig.~\ref{fig4}(a) and (b), the trace distance decays during the time evolution for both strong and weak thermalization, indicating the tendency $\hat{\rho}_{t}^{j,j+1} \simeq \hat{\rho}_{\beta}^{j,j+1}$. However, for weak thermalization, the trace distance strongly oscillates [Fig.~\ref{fig4}(b)].

Finally, we experimentally investigate the concurrence of the two-qubit reduced density matrix $\hat{\rho}_{t}^{j,j+1}$, which is defined as $E(\hat{\rho}) = \max\{0,\sqrt{\gamma_{1}}-\sqrt{\gamma_{2}} -\sqrt{\gamma_{3}} -\sqrt{\gamma_{4}} \}$, where $\gamma_{1}$, ..., $\gamma_{4}$ are eigenvalues listed in decreasing order of the matrix $\Gamma = \hat{\rho}(\hat{\sigma}^{y}\otimes \hat{\sigma}^{y})\hat{\rho}^{*}(\hat{\sigma}^{y}\otimes \hat{\sigma}^{y})$~\cite{concurrence1}. The time evolution of the concurrence, with the initial state $|\pi,0\rangle$ and $|\pi/2,\pi/4\rangle$, are presented in Fig.~\ref{fig4}(c) and (d), respectively. We observe that the concurrence vanishes after $t\simeq 50$ ns for strong thermalization. Whereas, the concurrence preserves a finite value with the initial state in the weak-thermalization regime, which can be interpreted as the thermal entanglement, i.e., the concurrence in thermal states $\hat{\rho}_{\beta}$~\cite{concurrence2}, according to $\hat{\rho}_{t}^{j,j+1} \simeq \hat{\rho}_{\beta}^{j,j+1}$ as a result of the ergodic dynamics in the weak-thermalization regime. The numerics of the concurrence in thermal states $\hat{\rho}_{\beta}$ with different $\beta$ are presented in Supplementary Materials~\cite{SM}.

In summary, we have provided a clear experimental evidence for characterizing the regimes of strong and weak thermalization. Weak thermalization, with a slow growth of the EE, has the potential for generating states with long-lived coherence and stabilizing the phases of matter far away from equilibrium, such as Floquet symmetry-protected topological phases~\cite{noneq_phase1}, discrete time crystals~\cite{noneq_phase2}, many-body localized phase~\cite{thermal5,mbl_sum1,mbl_sum2}, and dynamical paramagnetic and ferromagnetic phases~\cite{sciadv,noneq_phase3}. Our work also indicates that the 1D array of superconducting qubits can be a promising platform for exploring the issues at the foundation of quantum thermodynamics.

\begin{acknowledgments}
\noindent The authors thank the USTC Center for Micro- and Nanoscale Research and Fabrication. The authors also thank QuantumCTek Co., Ltd. for supporting the fabrication and the maintenance of room temperature electronics. This research was supported by the National Key R$\&$D Program of China (Grants No. 2018YFA0306703, No. 2017YFA0304300, No. 2016YFA0302104, No. 2016YFA0300600), the Chinese Academy of Sciences, and Shanghai Municipal Science and Technology Major Project (Grant No. 2019SHZDZX01), the Strategic Priority Research Program of Chinese Academy of Sciences (Grant No. XDB28000000), Japan Society for the Promotion of Science (JSPS) Postdoctoral Fellowship (Grant No. P19326), JSPS KAKENHI (Grant No. JP19F19326), the National Natural Science Foundation of China (Grants No. 11574380, No. 11905217, No. 11934018, No. 11774406), the Key-Area Research and Development Program of Guangdong Province (Grant No. 2020B0303030001), and Anhui Initiative in Quantum Information Technologies.
\end{acknowledgments}

%\noindent \textbf{Competing interests:} The authors declare no competing interests.

%\noindent \textbf{Data availability:} All relevant data are available from the corresponding authors upon request.

%\noindent \textbf{Author contributions:} %H.F., X.Z., and J.-W.P. conceived the research. Q.Z., Z.-H.S., M.G. and X.Z. designed the experiment. Q.Z. designed the sample. Q.Z., H.D., and H.R. prepared the sample. Q.Z. and M.G. carried out the measurements. Y.W. developed the programming platform for measurements. Z.-H.S. and C.Z. did numerical simulations. Q.Z., Z.-H.S., M.G., F.C., Y.-R.Z. and Y.Y. analyzed the results. Q.Z., Z.-H.S., M.G., Y.-R.Z., H.F., X.Z. co-wrote the manuscript. J.L., Y.X., L.S., C.G., F.L., and C.-Z.P. developed room temperature electronics equipments. All authors contributed to discussions of the results and development of manuscript. X.Z. and J.-W.P. supervised the whole project.

%\bibliography{reference_chen}

%%%%%%%%%%%%%%%%%%%%%%%%%%%%%%%

%%%%%%%%%%%%%%%%%%%%%%%%%%%%%%%
\clearpage
\renewcommand\thefigure{S\arabic{figure}}
\setcounter{figure}{0}  
\renewcommand\thetable{S\arabic{table}}

\noindent \emph{\textbf{Supplementary Materials for `Observation of strong and weak thermalization in a superconducting quantum processor'}}

\section{Device information and system Hamiltonian }

The experiment is performed on a ladder-type superconducting circuit consisting of 24 transmon qubits, which is the same circuit presented in Ref.~[1]. One can see more details about the device, the readout and the qubit manipulation in Tab.~\ref{table1} or Ref.~[1]. As shown in the Fig. 1(a) in the main text, the qubits $Q_{1}-Q_{12}$ are employed to the quantum simulation of an one-dimensional system.

%The optical micrograph of the 24-qubit superconducting circuit is presented in Fig.~\ref{figs0}\textbf{a}. Each qubit has an on-chip control line, enabling both tunning the qubit frequency (Z control) and applying the microwave drive (XY control). Moreover, each qubit has a dispersively coupled resonator for the state readout. One can see Fig.~\ref{figs0}\textbf{b} for more details.

\begin{table*}[t]
	\centering
	\scriptsize
	\renewcommand\arraystretch{1.8}
	\resizebox{\textwidth}{!}{
		\begin{tabular}{c |c c c c c c c c c c c c c}
			\hline
			\hline
			
			&$\text{Q}_{1}$ &$\text{Q}_{2}$ &$\text{Q}_{3}$ &$\text{Q}_{4}$ &$\text{Q}_{5}$ &$\text{Q}_{6}$			
			&$\text{Q}_{7}$ &$\text{Q}_{8}$ &$\text{Q}_{9}$ &$\text{Q}_{10}$ &$\text{Q}_{11}$ &$\text{Q}_{12}$\\
			\hline
			$\omega_{\textrm{read}}/2\pi$~(GHz)
			&6.688	&6.729 &6.790	&6.832	&6.886	&6.927	&6.701	&6.754	&6.801	&6.855	&6.909	&6.954\\
			$\omega_{\textrm{q}}^{\textrm{max}}/2\pi$~(GHz)
			&4.928  &5.536	&4.962  &5.600  &4.887	&5.600  &4.941	&5.562	&4.904	&5.602	&4.905  &5.587\\
			$\omega_{\textrm{q}}^{\textrm{idle}}/2\pi $~(GHz)
			&4.835  &5.31	&4.693	&5.38	&4.82	&5.26	&4.68	&5.32	&4.74	&5.27	&4.8	&5.4\\
			$T_1$~($\mu$s)
			&20.6  	&24.3	&28.1	&24.0	&26.3	&27.3	&21.4	&25.4	&21.4	&18.8	&22.2	&18.7\\
			$T_2^*$~($\mu$s)
			&3.9    &2.2	&2.0    &2.0	&4.9    &1.8    &2.1    &2.0    &2.3	&1.9    &4.4	&2.8\\		
			$U/2\pi $~(MHz)
			&$-238$ &$-230$ &$-240$ &$-230$ &$-238$  &$-230$ &$-240$ &$-230$ &$-236$ &$-230$&$-236$ &$-230$\\
			$J/2\pi $~(MHz)
			&\multicolumn{12}{c}{~~~12.4~~~~~~12.3~~~~~12.4~~~~~12.4~~~~~12.2~~~~~13.3~~~~~13.6~~~~~13.7~~~~~13.8~~~~~13.7~~~~~13.6~~~} \\
			$f_{00}$~($\textrm{\%}$)
			&97.6   &98.7	&98.2   &98.7	&98.9  &97.9   &98.9    &98.9	&98.6    &99.1	&97.8   &96.7\\
			$f_{11}$~($\textrm{\%}$)
			&91.2   &94.2	&94.1   &91.6	&93.6  &88.5   &93.8   &94.8    &92.3	&95.6    &94.1	&92.6\\
			
			\hline
			$1$Q RB fidelity~($\textrm{\%}$)
			&99.86 &99.88 &99.87 &99.88 &99.93 &99.91 &99.94 &99.87 &99.74 &99.94 &99.94 &99.91\\			
			\hline

		\end{tabular}	
	}
	
	\caption{ Qubit performance.
		{Parameters of the device. $w_{\text{read}}/2\pi$ is the resonant frequency of the readout resonator; $w_q^{\text{max}}$ is the maximum frequency of the qubit; $w_q^{\text{idle}}$ is qubit's idle frequency; $T_1$ and $T_2^*$ are the energy relaxation time and Ramsey dephasing time measured at the idle frequency; $U$ is the anharmonicity of qubit. $J/2\pi $ is the coupling strength of the corresponding qubit-pair measured at the working frequency ($4.863$~GHz). $1$Q RB fidelity is the average gate fidelity of single-qubit $X/2$ gate measured at idle frequency. The gate time is 30 ns.}
	}
	\label{table1}	
\end{table*}

Without the external transverse field, the Hamiltonian of the one-dimensional (1D) array of superconducting qubits can be written as
\begin{eqnarray} \nonumber
\hat{H} &=& \sum_{j=1}^{11} J_{j,j+1} (\hat{a}_{j}^{\dagger}\hat{a}_{j+1}+ \text{H.c.}) \\
&+&  \sum_{j=1}^{12}\frac{U_{j}}{2}\hat{n}_{j}(\hat{n}_{j}-1) +\sum_{j=1}^{12}\mu_{j}\hat{n}_{j}.
\label{SQP_BH}
\end{eqnarray}
The values of the hopping interaction between the $j-$th and $(j+1)-$th qubit, i.e., $J_{j,j+1}$ are presented in the Fig. 1(a) in the main text and Tab.~\ref{table1}, with an average value $\overline{J}/2\pi\simeq 12.3$ MHz. The average value of the onset nonlinear interactions $U_{j}$ is $\overline{U}/2\pi\simeq -234.0$ MHz. Because of $|\frac{\overline{U}}{\overline{J}}|\simeq 19$, the Bose-Hubbard model (\ref{SQP_BH}) is quite close to the hard-core limit where the bosonic creation and annihilation operator are mapped to the spin raising and lowering operator. When all qubits are tuned to the same working point, the Hamiltonian with the hard-core limit can be written as
\begin{eqnarray} \nonumber
\hat{H}_{XX} &=&  \sum_{j=1}^{11} J_{j,j+1} (\hat{\sigma}_{j}^{+}\hat{\sigma}_{j+1}^{-} + \text{H.c.}) \\
&\simeq& \lambda \sum_{j=1}^{11} (\hat{\sigma}_{j}^{x}\hat{\sigma}_{j+1}^{x} + \hat{\sigma}_{j}^{y}\hat{\sigma}_{j+1}^{y})
\label{XY1}
\end{eqnarray}
with $\lambda = \overline{J}/2$. By employing the Jordan-Wigner transformation $\hat{\sigma}_{j}^{-}=\exp(-i\pi\sum_{m<j}\hat{c}_{m}^{\dagger}\hat{c}_{m})\hat{c}_{j}$ and the Fourier transformation $\hat{d}_{k} = \frac{1}{\sqrt{N}} \sum_{j=1}^{N}\sum_{j=1}^{N}\exp(-ijk)\hat{c}_{j}$, the Hamiltonian can be rewritten as
$\hat{H}=\sum_{k} J\cos k \hat{d}_{k}^{\dagger}\hat{d}_{k}$ with the momentum $k$ taking $N$ (the number of qubit) discrete values from the Brillouin zone $k = \frac{2\pi n}{N}$ with $n\in\{-N/2+1,...,N/2\}$. Consequently, a 1D array of superconducting qubits can be regarded as an integrable system described by free fermions~[3].

%\begin{figure*}[]
%	\centering
%	\includegraphics[width=0.9\linewidth]{supp_Fig0_new.pdf}\\
%	\caption{\textbf{The 24-qubit superconducting circuit.} \textbf{a}, The optical micrograph of the 24-qubit superconducting circuit. The circuit is arranged into two rows of 12 qubits, and we employ the 12 qubits in the upper row of the circuit to perform the experiment. \textbf{b}, The details of the part of the device marked by the red frame in \textbf{a}. The qubit, XY control line, Z control line, bandpass filter, readout resonator, and readout line are highlighted. }\label{figs0}
%\end{figure*}

Numerical works have shown that strong and weak thermalization occur in 1D non-integrable systems. To observe strong and weak thermalization, we apply resonant microwaves to each qubit, which induces a local transverse field described by~[2]
\begin{eqnarray}
\hat{H}_{\text{drive}} =  \sum_{j=1}^{12} g_{j} (e^{-i\varphi_{j}} \hat{\sigma}^{+}_{j} + e^{i\varphi_{j}} \hat{\sigma}^{-}_{j})
\label{drive}
\end{eqnarray}
with $g_{j}$ and $\varphi_{j}$ as the strength and phase of the local transverse field imposed on the $j-$th qubit. In our experiment, we adopt that $g/2\pi=g_{j}/2\pi = 6$ MHz, and $\varphi=\varphi_{j}=\pi/2$. Finally, the Hamiltonian of the qubit chain with a local transverse field reads
\begin{eqnarray} \nonumber
\hat{H} &=&  \hat{H}_{XX} + \hat{H}_{\text{drive}} \\
&=& \lambda \sum_{j=1}^{11} (\hat{\sigma}_{j}^{x}\hat{\sigma}_{j+1}^{x} + \hat{\sigma}_{j}^{y}\hat{\sigma}_{j+1}^{y}) + g\sum_{j=1}^{12} \hat{\sigma}_{j}^{y}.
\label{SQP_final}
\end{eqnarray}

It is recognized that when we tune all qubits to the working point employing Z pluses, there is a disorder in the chemical potential, i.e., $\hat{H}_{\text{disorder}} = \sum_{j=1}^{12}\mu_{j}\hat{\sigma}_{j}^{+}\hat{\sigma}_{j}^{-}$, due to the crosstalk of the Z pluses. Here, we estimate that $\mu_{j}/2\pi \in [-1,1]$ MHz. In addition, the crosstalk of the local transverse field should also be considered. The strength of the local transverse field $g_{j}$ would satisfy $g_{j}/2\pi=6.7 \pm 1$ MHz (see below comparison between the numerics and experimental data).

\begin{figure}[]
	\centering
	\includegraphics[width=1\linewidth]{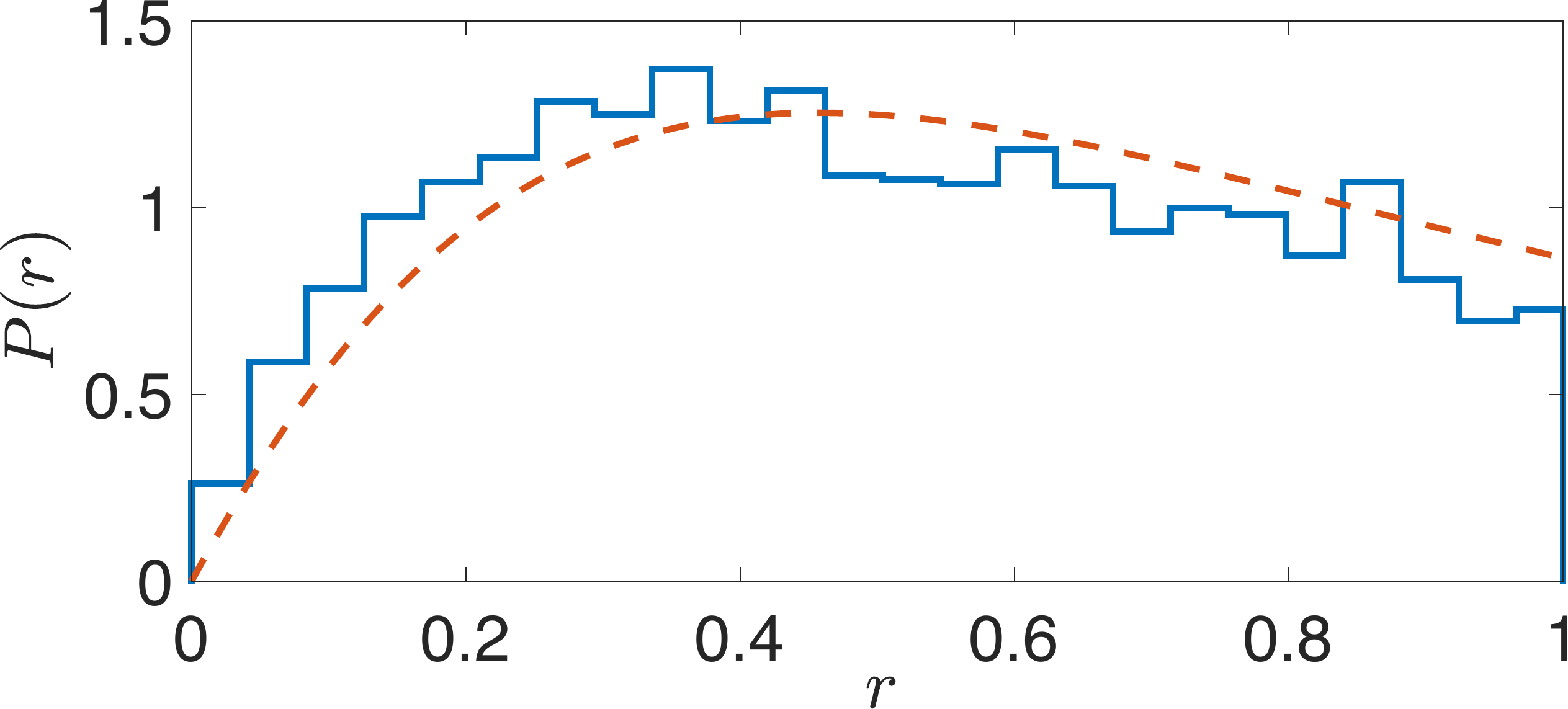}\\
	\caption{\textbf{Level statistics of the system Hamiltonian.} The dashed line is the probability distribution of $r_{n}$ following the GOE. The solid line is the numerics of $P(r)$ of the system (\ref{SQP_final}). }\label{figs_a1}
\end{figure}

We then discuss the integrability breaking originated from the local transverse field. Without the local transverse field, in the system (\ref{XY1}), the total number $\sum_{j=1}^{12} \hat{\sigma}_{j}^{+}\hat{\sigma}_{j}^{-}$ is conserved. The local transverse field can break the conservation. Moreover, we can study the level spacing distribution to characterize the non-integrability of the system with the local transverse field. Let $E_{n}$ as the eigenenergy of a Hamiltonian, and the level separation uniformity is defined as
\begin{eqnarray}
r_{n} = \frac{\min\{s_{n},s_{n-1}\}}{\max\{s_{n},s_{n-1}\}},
\label{rog}
\end{eqnarray}
where $s_{n} = E_{n+1} - E_{n}$ is the nearest-neighbor spacings.
It has been postulated that for a non-integrable system in the ergodic phase, the statistics of energy levels follows the Gaussian orthogonal ensemble (GOE)~[4], i.e.,
\begin{eqnarray}
P_{\text{GOE}}(r) = \frac{27}{4}\frac{r + r^{2}}{(1+r+r^{2})^{5/2}}.
\label{GOE}
\end{eqnarray}
As shown in Fig.~\ref{figs_a1}, the level statistics of the system Hamiltonian (\ref{SQP_final}) follows the GOE and the superconducting qubit chain with the local transverse field is a non-integrable system.

\section{Calibration}
In this experiment, in order to realize a non-integrable system where thermalization occurs, there are three requirements we have to meet: (1) All qubits should be tuned to the same interacting frequency; (2) The initial state of all qubits should be uniform at the start point of the time evolution; (3) The local transverse field of all qubits should be unifrom. We use the multi-qubit excitation propagation to calibrate frequency alignment of all qubits~[1], meeting the requirement (1). In the following section, we explain how to calibrate the initial states [the requirement (2)] and the transverse fields [the requirement (3)].

\subsection{Calibration of the initial state}
\begin{figure}[]
	\centering
	\includegraphics[width=1\linewidth]{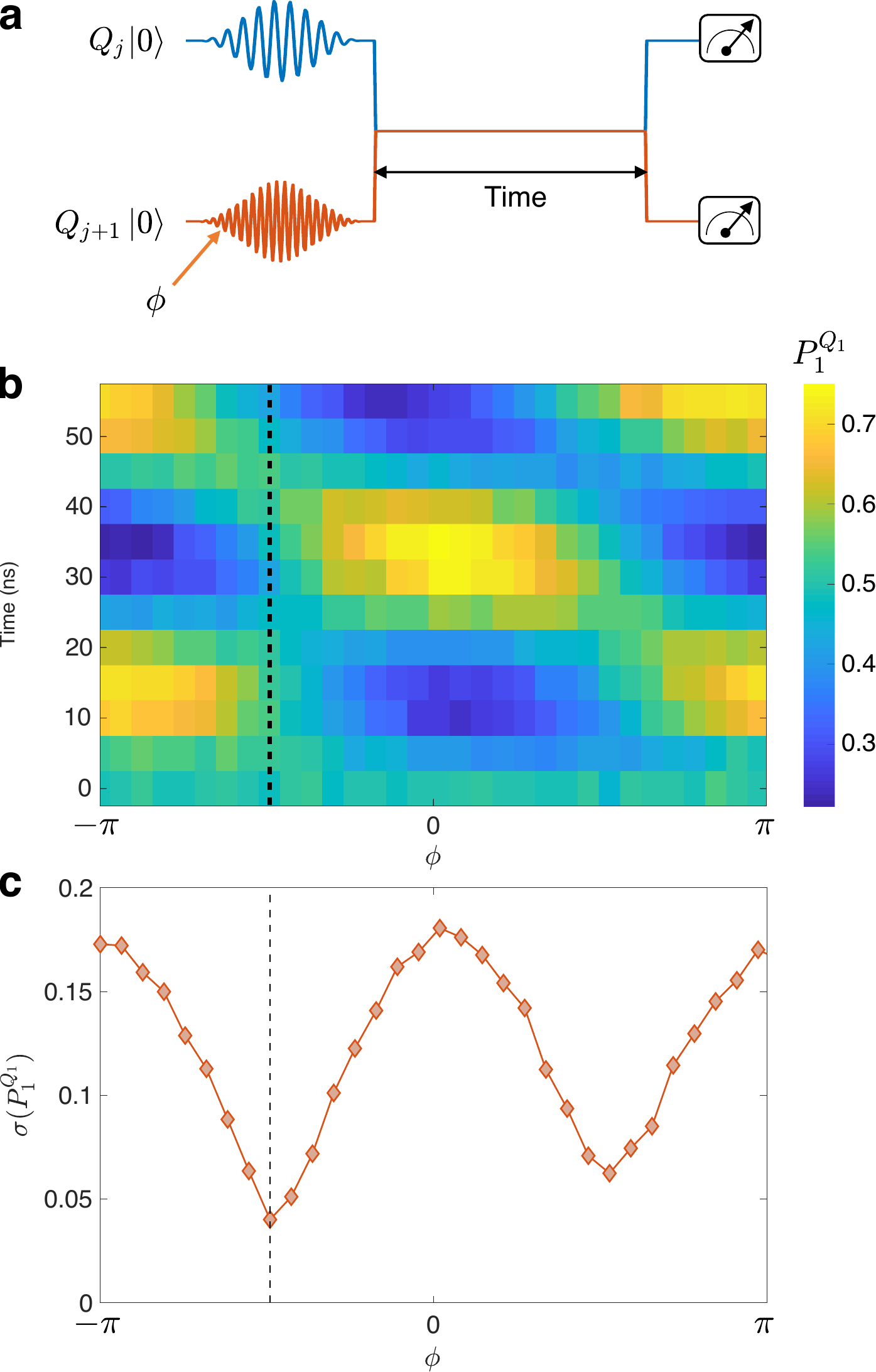}\\
	\caption{\textbf{Phase alignment of the rotation pulse employed to the initialization.} \textbf{a,} Experimental sequence for phase alignment of the initial state. \textbf{b,} The experimental result for phase
		alignment of the initial state. We plot the measured excited probabilities $P_{1}^{Q_{1}}$ as a function of the phase $\phi$ and the evolution time $t$. \textbf{c}, The standard deviation of $P_{1}^{Q_{1}}$ with different $\phi$. The dashed lines in \textbf{b} and \textbf{c} refer to the phase $\phi$ with minimum standard deviation of $P_{1}^{Q_{1}}$.  }\label{cali_1}
\end{figure}

\begin{figure}[]
	\centering
	\includegraphics[width=1\linewidth]{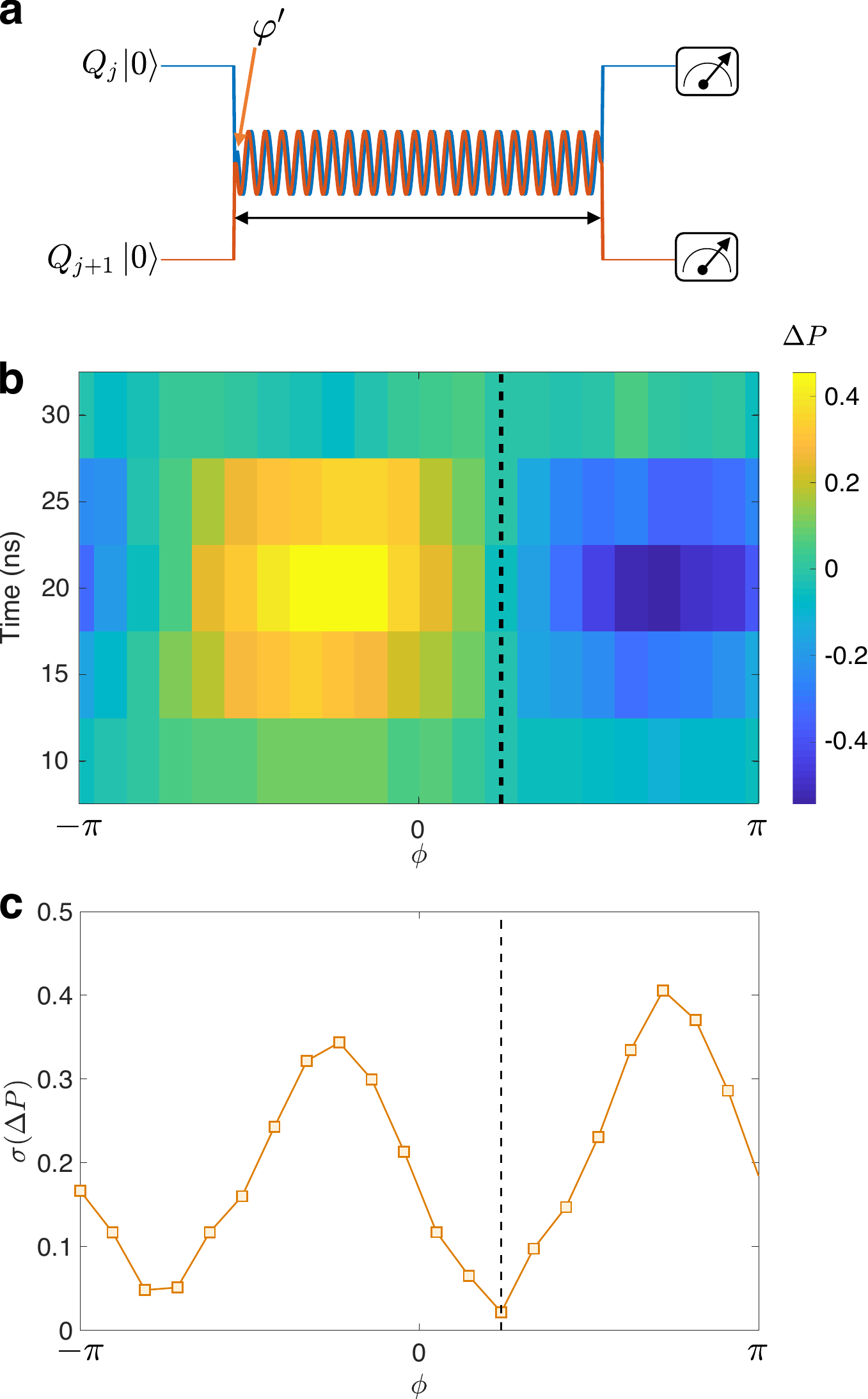}\\
	\caption{\textbf{Phase alignment of the transverse field} \textbf{a,} Experimental sequence for phase alignment of the transverse field.  \textbf{b,} The experimental result for phase alignment of the transverse field. We plot the measured difference of probabilities  $\Delta P_{1} = P_{1}^{Q_{1}} - P_{2}^{Q_{2}}$ as a function of the phase $\phi$ and the evolution time $t$. \textbf{c}, The standard deviation of $\Delta P_{1}$ with different $\phi$. The dashed lines in \textbf{b} and \textbf{c} refer to the phase $\phi$ with minimum standard deviation of $\Delta P_{1}$.  }\label{cali_2}
\end{figure}

In our experiment, the initial state of each qubit is prepared using a single qubit gate $\hat{R}(\theta_{0},\phi_{0})$, which is realized by resonant microwave pulses. It is required that all 12 qubits are in the same initial state at the start point of the time evolution, as shown
in the Fig. 1(b) and (c) of the main text. Therefore, for each qubit, both the $\theta_{0}$ and $\phi_{0}$ need extra calibrations. The $\theta_{0}$ is determined by the amplitude and length of the
microwave pulse. We can perform Rabi oscillation measurements on each qubit at the idle frequencies to ensure the uniformity of the $\theta_{0}$. Moreover, the $\phi_{0}$ is determined by the phase of the microwave pulse. As the qubits will accumulate extra dynamical phases
during the process tuned to the working frequency, we still need to calibrate the overall phases before the time evolution.

The main idea of the calibration is that when two nearest-neighboring qubits are tuned in resonance, if both $\theta_{0}$ and $\phi_{0}$ are uniform, there will not be any swapping between them. The calibration process of the initial state is shown in Fig.~\ref{cali_1}\textbf{a}. We take the qubit $Q_{1}$ and $Q_{2}$ as an example. We select the phase of the microwave pulse imposed on $Q_{1}$ as the frame of reference. We firstly apply an $X_{\pi/2}$ rotation pulse to $Q_{1}$ and a $\pi/2$ pulse whose rotation axis is of phase $\phi '$ in the $x$-$y$ plane to $Q_{2}$. Then, we bias $Q_{1}$ and $Q_{2}$ to the working frequency $\omega_{I}$ with a rectangular pulse. After a period of evolution, we bias them to their idle frequencies $\omega_{\text{q}}^{\text{idle}}$ and measure the population of state $|1\rangle$, i.e., $P_{1}^{Q_{1}}$. When
the initial states of $Q_{1}$ and $Q_{2}$ are the same, there will not be any swapping and the measured $P_{1}^{Q_{1}}$ will not oscillate during the dynamics.

The experimental results of $P_{1}^{Q_{1}}$ are displayed in Fig.~\ref{cali_1}\textbf{b}. To extract the phase $\phi$ with the weakest oscillation, we also study the standard deviation of $P_{1}^{Q_{1}}$ with respect to the evolution time $t$, i.e., $\sigma(P_{1}^{Q_{1}})$. The $\sigma(P_{1}^{Q_{1}})$ as a function of $\phi$ is displayed in Fig.~\ref{cali_1}\textbf{c}. The phase used for the correction is marked by the black dashed vertical line. We calibrate all nearest-neighboring qubits pairs with the same method. After that, all qubits are prepared to the same initial state before the time evolution.

\subsection{Calibration of the transverse field}

The experiment requires the uniformity of the local transverse field for each qubit, which means the $g_{j}$ and $\varphi_{j}$ in Eq.~(\ref{drive}) should be the same for all qubits. The parameters $g_{j}$ is determined by the amplitude of microwave pulse $A_{j}$.

There are two factors we need to consider in calibration: (1) The difference of the microwave phases caused by the length difference of qubits¡¯ control lines. (2) The XY-crosstalk caused by the microwave leakage. Our goal is to get 12 pairs of microwave pulse parameters $(A_{j},\varphi_{j}')$, which ensure the uniformity of all transverse fields. The initial value of $\varphi_{j}'$ is set to 0. We use the following procedure to calibrate the transverse field parameters.

1. By tuning each qubit $Q_{j}$ to the interacting frequency and performing Rabi oscillation measurements, we can get the initial values of the driving amplitude $A_{j}$ $(j=1,2,...,12)$ with $\Omega_{j}/2\pi=6$ MHz.

2. We use sequence in Fig.~\ref{cali_2}\textbf{a} to calibrate the phase of transverse field. Taking $Q_{1}$ and $Q_{2}$ as an example. We start with
initializing them in the ground state, then bias them to the interacting frequency, and apply resonant microwave drives on them
with the same magnitude but a phase difference of $\varphi_{j}'$. Note that when driving $Q_{1}$ and $Q_{2}$ , we simultaneously drive other qubits
to realize the effect of XY-crosstalk from other qubits. After a period of evolution, we bias $Q_{1}$ and $Q_{2}$ to their idle frequencies
and measure their population of $|1\rangle$. When the phases of the transverse fields are aligned, the difference between the population of $Q_{1}$ and $Q_{2}$, which is defined as $\Delta P_{1} = P_{1}^{Q_{1}} - P_{2}^{Q_{2}}$, will not oscillate. The experimental results of $\Delta P_{1}$ are displayed in Fig.~\ref{cali_2}\textbf{b}. The standard deviation of $\Delta P_{1}$ with respect to the evolution time $t$, i.e., $\sigma(\Delta P_{1})$, is plotted in Fig.~\ref{cali_2}\textbf{c}. We perform this calibration to all nearest-neighboring qubits pairs. Then, the microwave phases $\varphi_{j}'$ are updated.

3. Again, we tune each qubit $Q_{j}$ to the interacting frequency and run Rabi oscillation measurements. There is a little difference
from step 1 to take the XY-crosstalk into account. When driving $Q_{j}$, we simultaneously apply microwave pulses to other qubits
which are in idle frequencies. Then we can update driving amplitudes $A_{j}$.

4. We iterate over step (2) and step (3) until the microwave pulse parameters $A_{j}$ and $\varphi_{j}'$ converge. Typically, we only need two
to three iterations.

\section{Numerics of the local observable and entanglement entropy}

\begin{figure}[]
	\centering
	\includegraphics[width=1\linewidth]{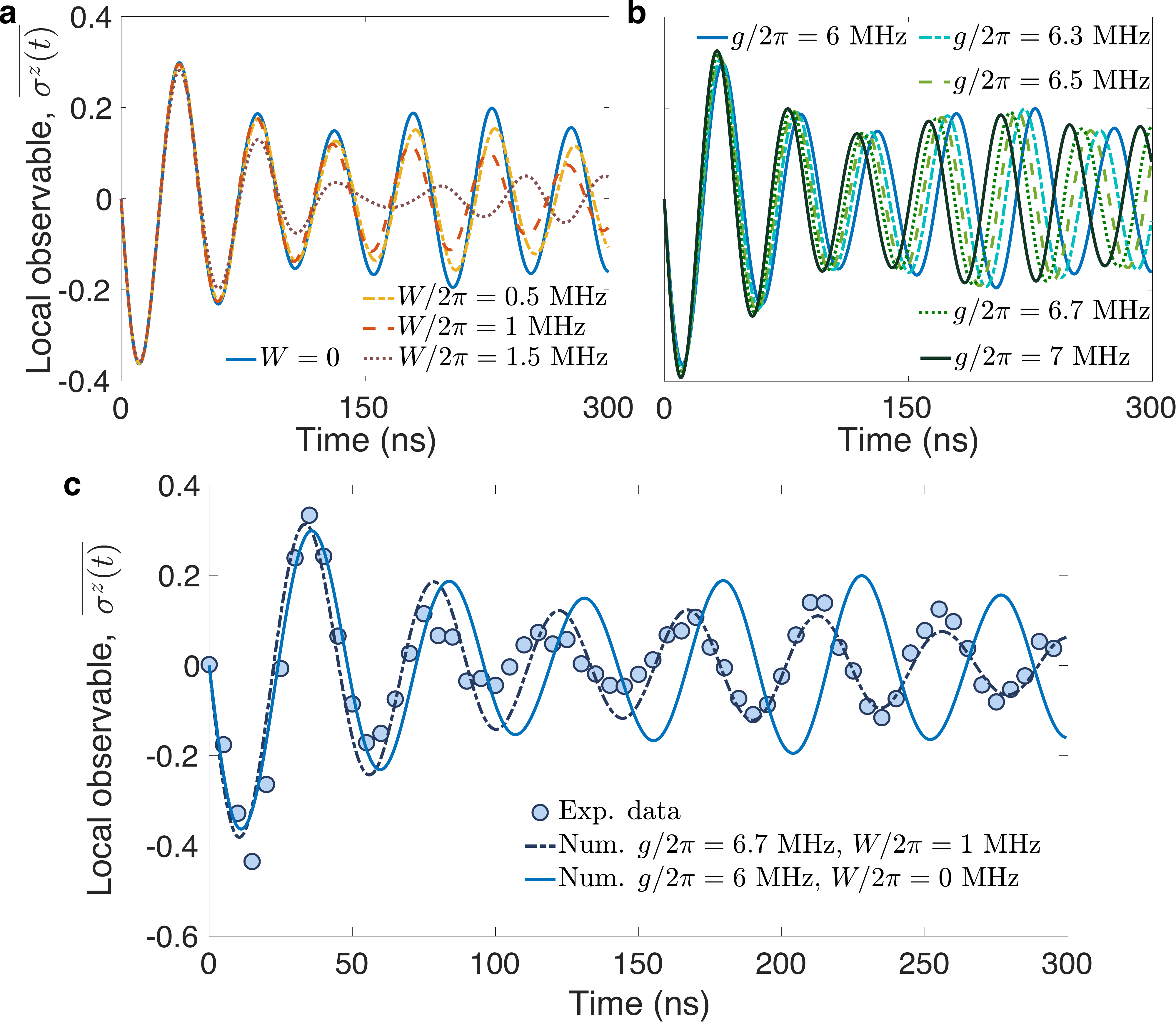}\\
	\caption{\textbf{Numerics of the local observable in comparison with the experimental data with the initial state $|\pi/2,\pi/4\rangle$.} \textbf{a,} Numerical data of the local observable with different disorder strength $W$. \textbf{b,} Numerical data of the local observable with different strength of the local transverse field $g/2\pi$. \textbf{c,} The comparison between the experimental and numerical data.  }\label{figs_num1}
\end{figure}

\begin{figure}[]
	\centering
	\includegraphics[width=1\linewidth]{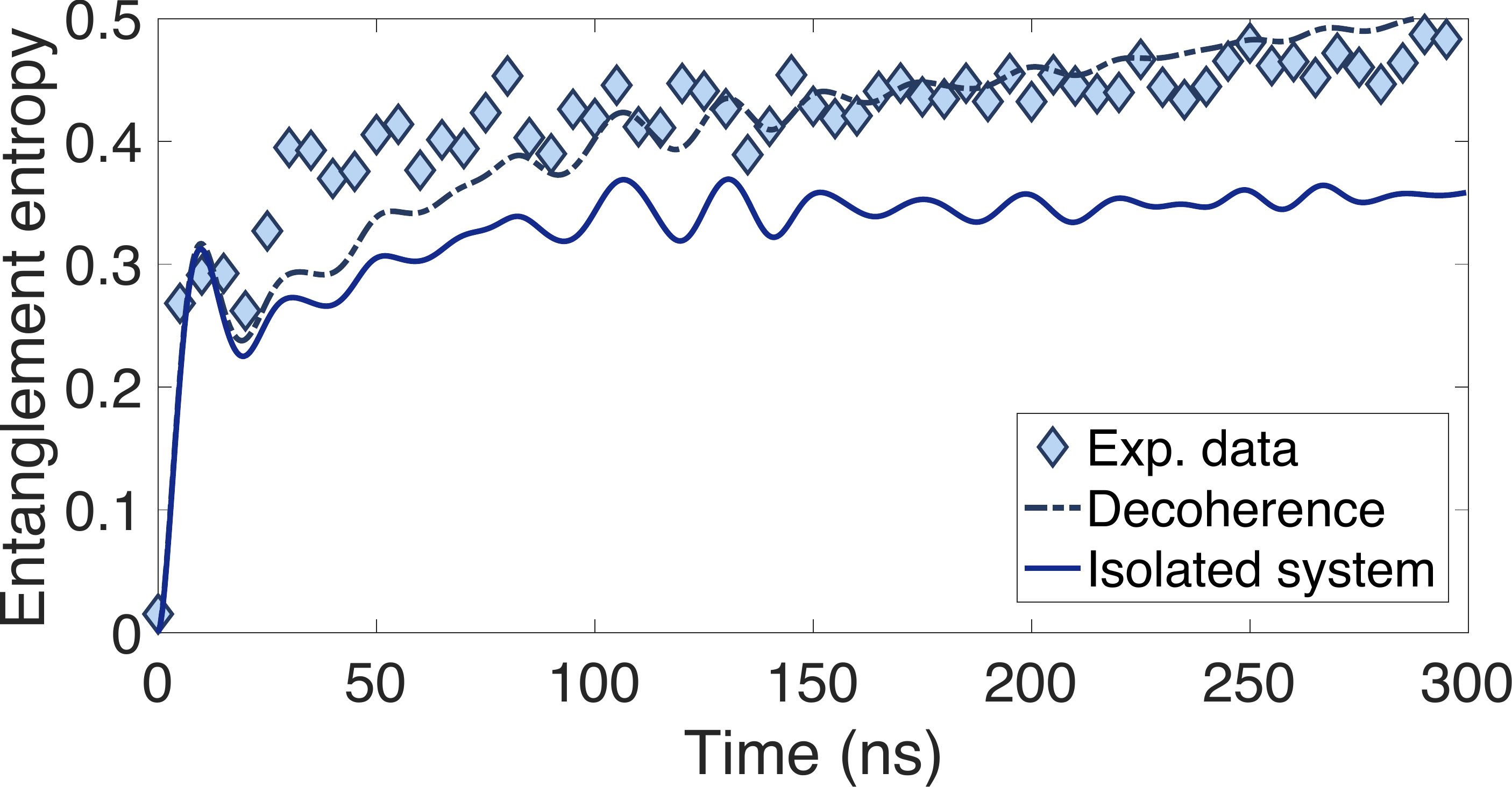}\\
	\caption{\textbf{The dynamics of the EE with the initial state $|\pi/2,\pi/4\rangle$.} The solid line is the numerical data obtained from unitary evolution. The dashed line is the numerical data considering decoherence effects. The diamond points are the experimental data. }\label{figs_num3}
\end{figure}

\begin{figure*}[]
	\centering
	\includegraphics[width=1\linewidth]{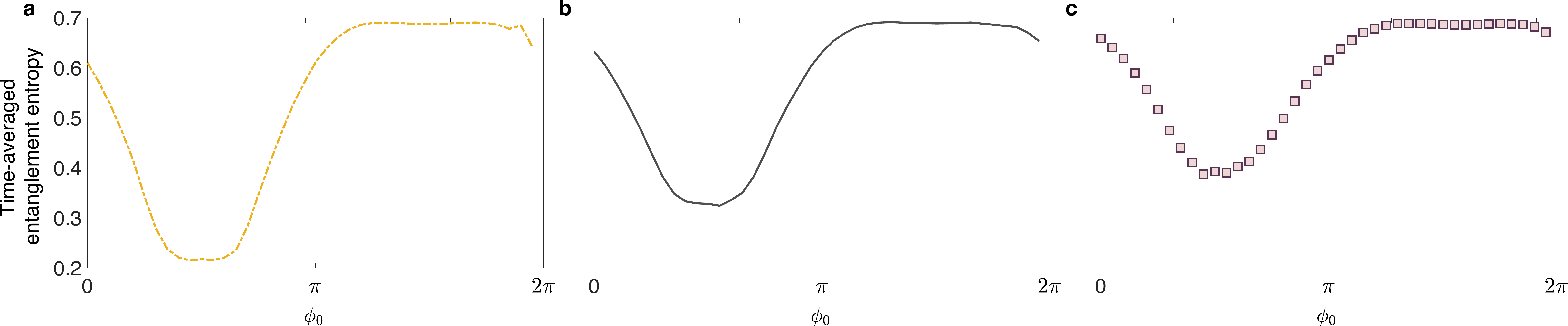}\\
	\caption{\textbf{The time-averaged EE.} \textbf{a,} Numerical data of the time-averaged EE for the initial states $|\pi/2,\phi_{0}\rangle$ ($\phi_{0}\in[0,2\pi]$), without decoherence, i.e., considering unitary evolution. \textbf{b,} Numerical data of the time-averaged EE for the initial states $|\pi/2,\phi_{0}\rangle$, taking the decoherence effects into consideration. \textbf{c,} Experimental data of the time-averaged EE for the initial states $|\pi/2,\phi_{0}\rangle$.}\label{figs_num6}
\end{figure*}

%\begin{figure}[]
%	\centering
%	\includegraphics[width=1\linewidth]{supp_num_2.pdf}\\
%	\caption{\textbf{Numerics of the local observable and EE in comparison with the experimental data for strong thermalization.} \textbf{a,} Numerics of the dynamics of the local observable with the initial state $|\pi,0\rangle$ in comparison with the experimental data. \textbf{b} is similar to \textbf{a} but with the initial state $|\pi/2,8\pi/5\rangle$. \textbf{c,} Numerics of the dynamics of the EE with the initial state $|\pi,0\rangle$ in comparison with the experimental data. \text{d} is similar to \textbf{c} but with the initial state $|\pi/2,8\pi/5\rangle$. The points and dashed lines refer to the experimental and numerical data, respectively.  }\label{figs_num2}
%\end{figure}

For the local transverse field $\hat{H}_{\text{driven}} = \sum_{j=1}^{12} g_{j}\hat{\sigma}_{j}^{y}$, we experimentally set $g/2\pi = g_{j}/2\pi = 6$ MHz. However, if we numerically simulate the dynamics of the local observable $\overline{\sigma^{z}(t)}$ using $g/2\pi = g_{j}/2\pi = 6$ MHz, there is an obvious discrepancy between the numerics and experimental data (see the solid line and circle points in Fig.~\ref{figs_num1}\textbf{c}). Thus, the crosstalk effect should be considered in detail.

We assume that the crosstalk can induce a disorder of the local transverse field, i.e., $g_{j}/2\pi \in (\overline{g} + [-W,W])/2\pi$ MHz with $\overline{g}/2\pi=6$ MHz and $W$ as the strength of disorder. The numerical results of the local observable with different $W$ is shown in Fig.~\ref{figs_num1}\textbf{a}. One can see that the amplitude of $\overline{\sigma^{z}(t)}$ is suppressed with the increase of $W$. Additionally, we assume that the strength of the local transverse field $\overline{g}/2\pi$ can be enlarged by the crosstalk effect. As shown in Fig.~\ref{figs_num1}\textbf{b}, the frequency of the oscillation of $\overline{\sigma^{z}(t)}$ is closely related to $\overline{g}/2\pi$. We consider the two parameters $\overline{g}$ and $W$, and find that when we chose $\overline{g}/2\pi=6.7$ MHz and $W/2\pi=1$ MHz, the numerics is in good agreement with the experimental data.

We also present the numerics of the local observable and entanglement entropy (EE) with the initial states in strong-thermalization regime [see the Fig. 2(a), (b), (d) and (e) in the main text]. The numerical simulations also consider $\overline{g}/2\pi=6.7$ MHz and $W/2\pi=1$ MHz and the results are consistent with the experimental data.

\section{The impact of decoherence on the entanglement entropy}

The above numerics mainly consider an isolated system evolved by the unitary transformation, i.e., $|\Psi_{t}\rangle = e^{-i\hat{H}t}|\theta_{0},\phi_{0}\rangle$. However, the decoherence is inevitable in experiments. The results in Fig.~\ref{figs_num1}\textbf{c} show a good agreement between the numerical and experimental data, indicating that the dynamics of local observable in the regimes of both strong and weak thermalization, and the dynamics of the EE for strong thermalization are robust against decoherence effects.

As shown in Fig.~\ref{figs_num3}, there exists an obvious difference between the numerics without considering decoherence and the experimental data. Consequently, we should numerically estimate the impacts of decoherence on the dynamics of the EE in the weak-thermalization regime. The dynamics with decoherence can be modeled by the Lindblad master equation [5,6]
\begin{eqnarray} \nonumber
\frac{\text{d} \hat{\rho}_{t}}{\text{d} t} = &-&i[\hat{H},\hat{\rho}_{t}] \\
&+& \frac{1}{2} \sum_{j=1}^{12}(2\hat{L}_{j}\hat{\rho}_{t}\hat{L}_{j}^{\dagger} - \{\hat{L}_{j}^{\dagger}\hat{L}_{j},\hat{\rho}_{t}\}  )
\label{lindblad}
\end{eqnarray}
with $\hat{L}_{j}$ ($\hat{L}_{j}^{\dagger}$) as the Lindblad operators and $\hat{\rho}_{t} = |\Psi_{t}\rangle\langle\Psi_{t} |$. There are two effects of decoherence: the energy relaxation and dephasing effect, identified by the energy lifetime $T_{1}$ and dephasing time $T_{2}$, respectively. The corresponding Lindblad operators are $\hat{L}_{j} = \hat{\sigma}_{j}^{-}/\sqrt{T_{1}}$ (energy relaxation effect) and $\hat{L}_{j} = \hat{\sigma}_{j}^{z}/\sqrt{2T_{2}}$ (dephasing effect). For our device, the averaged energy lifetime is $\overline{T_{1}}\simeq 23.6$ $\mu$s, and the averaged dephasing time $\overline{T_{2}}\simeq3.82$ $\mu$s. The numerical results of the dynamics of EE with decoherence effects displayed in Fig~\ref{figs_num3} are obtained by solving Eq.~(\ref{lindblad}) with $\overline{T_{1}}\simeq 23.6$ $\mu$s and $\overline{T_{2}}\simeq3.82$ $\mu$s. It is shown that the numerics with decoherence effects are consistent with the experimental data. The numerics of time-averaged entanglement entropy without and with considering the decoherence effects are presented in Fig~\ref{figs_num6}\textbf{a} and \textbf{b}, respectively. In comparison with the experimental data in Fig~\ref{figs_num6}\textbf{c}, the numerical data, taking the decoherence effects into consideration, are in better agreement with the experimental data.

\section{Numerics of the concurrence}

\begin{figure}[]
	\centering
	\includegraphics[width=1\linewidth]{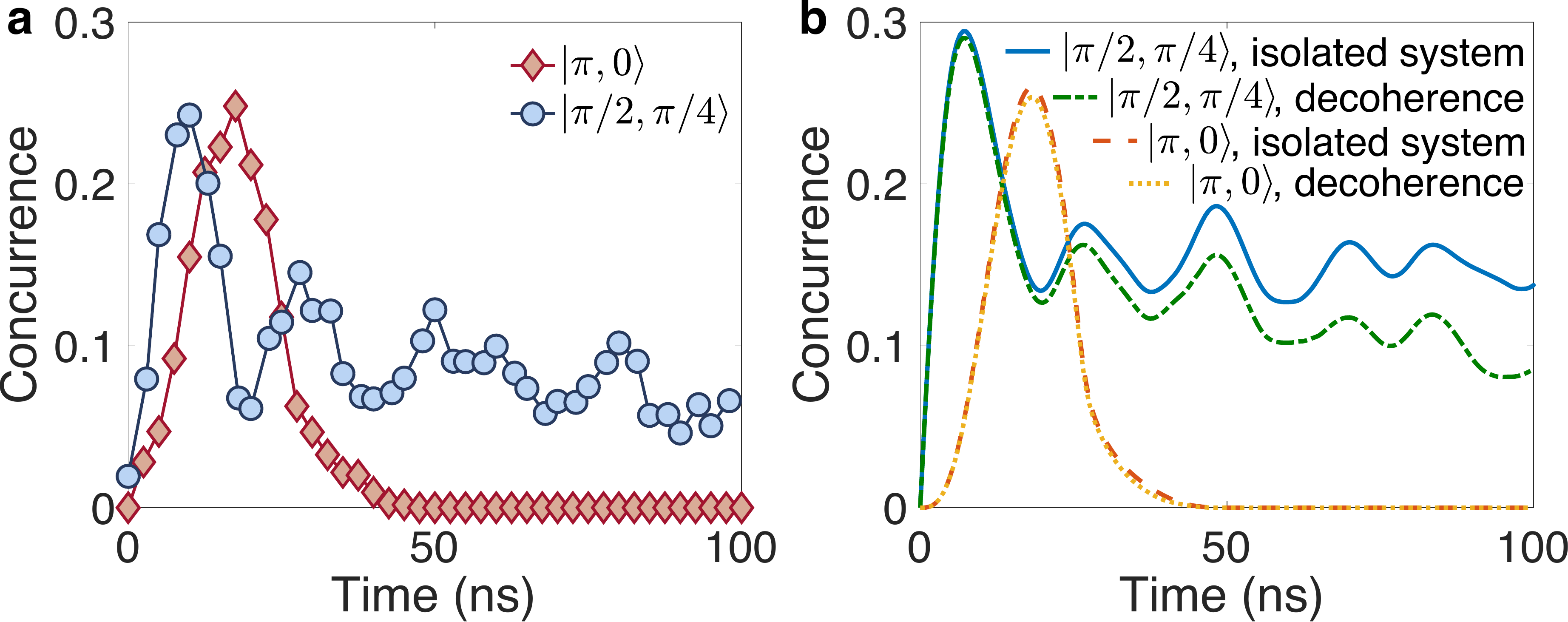}\\
	\caption{\textbf{Numerics of the concurrence in comparison with the experimental data.} \textbf{a,} Experimental data of the  concurrence with different initial states. \textbf{b,} Numerical results of the concurrence with and without considering decoherence effects.  }\label{figs_num4}
\end{figure}

\begin{figure}[]
	\centering
	\includegraphics[width=1\linewidth]{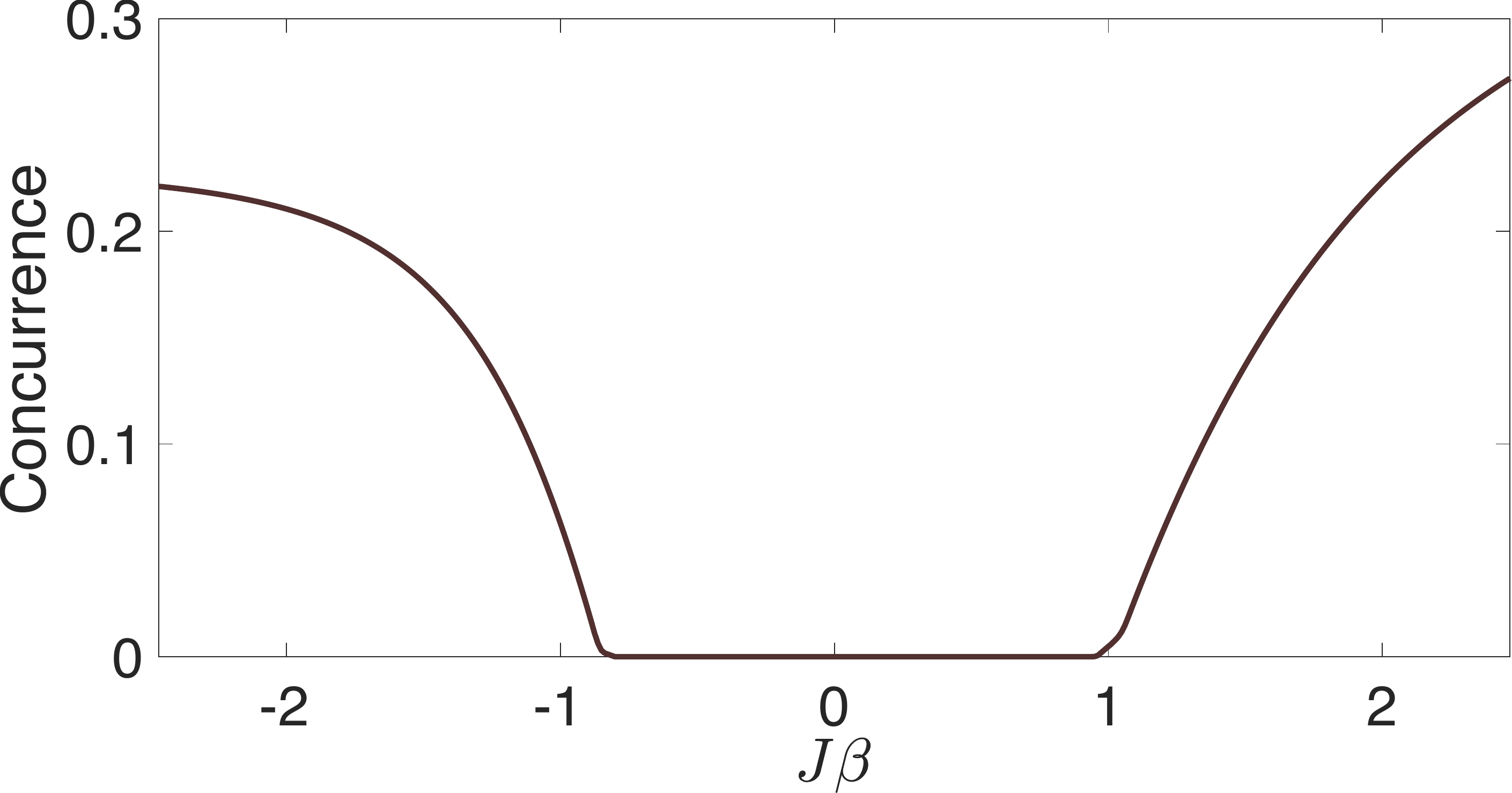}\\
	\caption{\textbf{Numerics of the thermal entanglement quantified by the concurrence.} The concurrence of thermal states $\hat{\rho}_{\beta}^{j,j+1}$, averaged over the qubit site $j$, as a function of the inverse temperature $\beta$. }\label{figs_num_a2}
\end{figure}

In Fig.~\ref{figs_num4}, we present the comparison between the numerical and experimental results of the concurrence. We show that the dynamics of concurrence with the initial state $|\pi/2,\pi/4\rangle$ in weak-thermalization regime is also apparently influenced by decoherence effects. We find an overall decrease of the concurrence as a function of time in the presence of decoherence effects.

In addition, we calculate the concurrence of the thermal state $\hat{\rho}_{\beta}^{j,j+1}\equiv\text{Tr}_{A\notin\{j,j+1\}} \hat{\rho}_{\beta}$ with $\hat{\rho}_{\beta}\equiv  \exp(-\beta\hat{H})/\text{Tr}[\exp(-\beta\hat{H})]$, and average the concurrence over the qubit site $j=1,2,...,11$. The averaged concurrence with different inverse temperature $\beta$ is displayed in Fig.~\ref{figs_num_a2}. As shown in Fig. 1 in the main text, the $\beta$ of the initial state $|\pi/2,\pi/4\rangle$ in the weak-thermalization regime is estimated as $J\beta\simeq -1.034$. In Fig.~\ref{figs_num_a2}, the concurrence of the thermal state $\hat{\rho}_{\beta}^{j,j+1}$ with $J\beta\simeq -1.034$, while the concurrence vanishes for the thermal state $\hat{\rho}_{\beta}^{j,j+1}$ with $-0.80\leq J\beta \leq 0.93$. Therefore, the finite value of the concurrence with the initial state $|\pi/2,\pi/4\rangle$, shown in Fig.~\ref{figs_num4}, can be interpreted as a thermal entanglement in the state $\hat{\rho}_{\beta}^{j,j+1}$ with $J\beta\simeq -1.034$, based on $\hat{\rho}_{t}^{j,j+1}\rightarrow \hat{\rho}_{\beta}^{j,j+1}$ (see the main text).

Next, we discuss the trace distance between the quenched states and the thermal states. For the quenched states, the reduced density matrices $\hat{\rho}_{t}^{j,j+1}\equiv\text{Tr}_{A\notin\{j,j+1\}} \hat{\rho}_{t}$ can be measured via the two-qubit quantum state tomography performed on the $j-$th and $(j+1)-$th qubit. Whereas, the thermal state $\hat{\rho}_{\beta}^{j,j+1}$ can only be numerically estimated. We emphasize that the thermal state $\hat{\rho}_{\beta}$ is dependent on the Hamiltonian $\hat{H}$. As we discussed in Section 2, there are random parameters in the Hamiltonian $\hat{H}$, i.e., $\hat{H}_{\text{drive}} = \sum_{j=1}^{12} (\overline{g} + \delta g_{j}) \hat{\sigma}_{j}^{y}$, where $\overline{g}/2\pi = 6.7$ MHz and $\delta g_{j}\in [-W,W]$ with $W/2\pi =1$ MHz.

For the initial state $|\pi,0\rangle$ in the strong-thermalization regime, the estimated inverse temperature of the initial state $|\pi,0\rangle$ is $\beta=0$ which is independent of the random parameters $\delta g_{j}$. Therefore, the thermal state is $\hat{I} = \frac{1}{4}\text{diag} (1,1,1,1)$. Nevertheless, for the initial state $|\pi/2,\pi/4\rangle$ in the weak-thermalization regime, the estimated inverse temperature $\beta$ is dependent on the random parameters. Here, we use 20 samples of $\delta g_{j}$ and estimate the $\beta$ of $|\pi/2,\pi/4\rangle$. Then we can calculate the thermal states $\hat{\rho}_{\beta}^{j,j+1}$ for the 20 Hamiltonian and average them over the 20 samples.

\indent 

%\begin{thebibliography}{9}
\noindent [1] Q. Zhu, \emph{et al.}, arXiv: 2101.08031.

\noindent [2] K. Xu, \emph{et al.}, Sci. Adv. \textbf{6}, eaba4935 (2020).

\noindent [3] E. Lieb, T. Schultz, and D. Mattis, Ann. Phys. \textbf{16}, 407 (1961).

\noindent [4] Y. Y. Atas, E. Bogomolny, O. Giraud, and G. Roux, Phys. Rev. Lett. \textbf{110}, 084101 (2013).

\noindent [5] R. Johansson, P. Nation, and F. Nori, Comput. Phys. Commun. \textbf{183}, 1760 (2011).

\noindent [6] R. Johansson, P. Nation, and F. Nori, Comput. Phys. Commun. \textbf{184}, 1234 (2013).

%\end{thebibliography}

\end{document}